\documentclass[twocolumn,amsmath,amssymb,reprint]{revtex4-2}
\usepackage{graphicx}
\usepackage{bm}
\usepackage{float}
\usepackage{color,soul}

\usepackage{subcaption}
\usepackage[sort&compress]{natbib}
\setcitestyle{maxnames=10}

\newcommand{\DefFigWidth}{.48\textwidth}

\begin{document}

\title{The interaction of turbulence, magnetic islands and zonal fields in fluid plasma models with cubic non-linearities}% Force breaks with \\

\author{D. Villa}
 \affiliation{Max Planck Institute for Plasma Physics, Boltzmannstraße 2, 85748, Garching, Germany}
 \affiliation{Excellence Cluster ORIGINS, Boltzmannstraße 2, D-85748, Garching, Germany}
 \email{daniele.villa@ipp.mpg.de}

\author{N. Dubuit}%
\affiliation{Aix-Marseille Universit\'e, CNRS, PIIM UMR 7345, Marseille, France}

\author{O. Agullo}
\affiliation{Aix-Marseille Universit\'e, CNRS, PIIM UMR 7345, Marseille, France}

\author{X. Garbet}
\affiliation{CEA, IRFM, F-13108 Saint-Paul-Lez-Durance, France}
\affiliation{School of Physical and Mathematical Sciences, Nanyang Technological University, 637371 Singapore}

\date{\today}

\begin{abstract}
It is shown that the generation of magnetic islands by pressure-gradient-driven turbulence is common across a wide range of conditions. The interaction {among the turbulence,} the magnetic island and other large scale structures, namely the zonal flow and the zonal current, largely determines the dynamics of the overall system. The turbulence takes a background role, providing energy to the large-scale structures, without influencing their evolution directly. It is found that the growth of the zonal current is linearly related to that of the magnetic island, while the zonal flow has a strongly sheared region where the island has its maximum radial extension. The zonal current is found to slow down the formation of large-scale magnetic islands, while the zonal flow is needed to have the system move its energy to larger and larger scales. The driving instability in the system is the fluid Kinetic Ballooning Mode (KBM) instability at high $\beta$, while the tearing mode is kept stable. The formation of magnetic-island-like structures at the spatial scale of the fluid KBM instability is observed quite early in the non-linear phase for most cases studied, and a slow coalescence process evolves the magnetic structures towards larger and larger scales. Cases that did not show this coalescence process, nor the formation of the small scale island-like structures, were seen to have narrower mode structures for comparable instability growth rates, which was achieved by varying the magnetic shear. The islands often end up exceeding the radial box size late in the non-linear phase, showing unbounded growth. The impact on the pressure profile of turbulence driven magnetic islands is not trivial, showing flattening of the pressure profile only far from the resonance, where the zonal flow is weaker, and the appearance of said flattening is slow, after the island has reached a sufficiently large size, when compared with collisional time scales.
\end{abstract}

\maketitle

\section{Introduction} 
Magnetic islands are a well-studied phenomenon in plasma physics. They are the result of a change in the topology of the magnetic field following a reconnection process. Since their formation is the result of a magnetic reconnection process, their presence can be associated, among other things, to localized heating and outflows, of particular interest, for example, in solar physics \citep{janvier2017three}. The interest of the fusion community in this phenomenon is mostly due to the negative effects they have on the confinement properties of the plasma, the instabilities they can trigger or drive and to the fact that they might lead to disruptions in tokamak discharges, thus potentially damaging the device they develop in.\\
Despite the very large volume of work done on the topic of magnetic islands, there are still some open questions about the fundamental properties of these structures. Of particular interest for this work is the generation of magnetic islands by turbulence \citep{yagi2007nonlinear, muraglia2011generation, agullo2017nonlinearI, agullo2017nonlinearII, dubuit2021turbislands} and the interaction of magnetic islands with other large-scale structures, such as zonal currents and zonal flows \citep{guzdar2001zonal,diamond2005zonal,fujisawa2007experimental,dong2019nonlinear}.\\
The mainstream experimental approach of using the Modified Rutherford Equation (MRE) to predict the instability and saturation of magnetic islands \citep{escande2004simple, militello2004simple} has been shown to have limited validity \citep{ishizawa2010turbulence, borgogno2014nonlinear, muraglia2021nonlinear}, thus more in depth studies are needed to address the stability, generation and non-linear dynamics of magnetic islands, especially when they are not driven by tearing mode instabilities.\\
In this paper, it will be shown how the generation of magnetic islands by turbulence is possible, and indeed consistently observed, for a range of systems with various plasma $\beta = \frac{2\mu_0 \; p}{B^2}$ (where $p$ is the plasma pressure, $B$ the magnetic field amplitude and $\mu_0$ the vacuum permeability constant), with varying magnetic shear and instability drive. In this work, continuation and extension of the results presented in \cite{villa2025zonal}, the formation of the island has an additional component with respect to the processes described in previous studies \citep{yagi2007nonlinear, muraglia2011generation, agullo2017nonlinearI, agullo2017nonlinearII, dubuit2021turbislands}, where the island was driven by direct coupling of the most unstable interchange mode $m^*$ with its neighbouring modes $m^* \pm 1$. Here this process is still present, but it is of secondary importance, as no magnetic island is observed in the system unless a change in parity of the dominant mode $m^*$ to tearing parity occurs (through coupling with the zonal field), which then adds to the sub-dominant magnetic islands formed through the $m^* \pm 1$ via an inverse cascade (in the broad sense of a transfer of energy from smaller to larger scales via non-linear interactions), and only then does a dynamically relevant island appear in the system.\\
{Furthermore}, the non-linear dynamics and transport properties of these so-called ``turbulence driven magnetic islands'' (TDMIs) are analysed, with particular focus on the interaction between these magnetic structures and the zonal fields, showing how the TDMI can drive sheared plasma flows at radial locations where its separatrix has the maximum radial extensions. It is also shown that the zonal flow is needed to have the formation of large-scale magnetic islands, while the zonal current slows down the formation of such structures. {This is a peculiarity of TDMIs, as opposed to magnetic islands driven by e.g. tearing modes}\cite{militello2004simple, escande2004simple, smolyakov2013higher, poye2013global, loizu2020direct}{ or micro-tearing modes }\cite{guttenfelder2011electromagnetic, giacomin2023nonlinear}{, as the latter are the direct result of a (non-)linear instability, that has its own well-defined growth properties and saturation mechanism, whereas the TDMI is generated, driven and saturated by other mechanisms, among which turbulence and the zonal fields, that are non-linear and don't require a specific background configuration to exist.}\\
With respect to the transport properties of the plasma, one of the most prominent effects of magnetic islands is that they have been shown both theoretically \citep{fitzpatrick1995helical} and experimentally \citep{snape2012influence} to cause a flattening of the radial pressure profile inside the separatrix (i.e. the surface that separates the island from the rest of the plasma) by connecting radially distinct positions through parallel streaming. This flattening is, to first approximation \citep{fitzpatrick1995helical}, only effective if the width of the island crosses a critical threshold in width. Beyond this threshold, the parallel diffusivity is more efficient than the perpendicular, and, without going into the details of further subtleties that arise from considerations of additional effects \citep{dudkovskaia2021drift, imada2018nonlinear, villa2022localized, tae2024unveiling}, it will then pose a problem for the performance of the plasma. A flat pressure gradient is also reason for concern for the growth of so-called Neoclassical Tearing Modes \citep{sauter1997beta} (NTMs), i.e. magnetic islands whose driving mechanism is the suppression of the bootstrap current linked to the fading pressure gradient \citep{carrera1986island, reimerdes2002current}. This kind of island requires the presence of a seed island above a critical size to become unstable, but can grow to sizes that are unsafe for the operation of magnetic confinement devices. This means that even if the magnetic configuration can be optimized to be stable against the main mechanisms that generate magnetic islands, any other process that could generate a seed island of sufficient width would still be problematic. Indeed, magnetic islands still show up even if the experimental setup doesn't directly explain their origin \citep{isayama2013onset}, which has implications for the experimental effort to reach higher plasma $\beta$. Recently their presence has also been detected at the pedestal of H-mode plasmas \citep{hoelzl2023non}, thus renewing interest in their influence over the confinement properties of fusion plasmas. {Here it will be shown how TDMIs can reach significant sizes in the system without having visible impact on the pressure profile, unless one waits significantly longer times than expected from collisional arguments, and how the resulting flattening of the pressure profile does not lie on the resonance but in the neighbouring regions.}\\
One might argue for the need to have realistic geometries and more sophisticated setups than used here, in order to assess the ``real-life'' impact of the phenomena investigated here. However, having the ability of studying a system where the role of the individual elements can be clearly identified is a necessary first step before diving into more complex situations, and is the focus of the present paper.\\
The properties of the fluid model employed in this study \citep{villa2022localized}, in particular in what relates to the so-called ``cubic non-linearities'' or ``cubic terms'' \citep{villa2022cubic} and their impact on the dynamics of TDMIs are addressed, and shown to lead to modifications of the properties of the system depending on whether or not they are considered in the simulations.\\
%The results presented in this paper only pertain to the generation of magnetic islands by turbulence, to properly assess the non-linear dynamics of the system multi-helicity 3D simulations will be necessary, so that the interaction of different resonances can be studied. The 2D simulations in this work allow to state that when turbulence reaches a magnetic surface with low safety factor the generation of magnetic islands on said surface becomes possible, and the main result presented here is that this can happen in a wide range of $\beta$ values.\\
The article develops as follows: in section \ref{sec_model} the fluid model used for the analysis is briefly described, in section \ref{sec_linear} the linear analysis of the system and some properties of the system related to the cubic terms are highlighted, in section \ref{sec_generation} results from 2D fluid simulations in slab geometry with regard to the generationof TDMIs are presented, followed by analysis of their dynamics in section \ref{sec_dynamics} and in section \ref{sec_discussion} some concluding remarks and an outlook for future development are expressed.

\section{The electromagnetic reduced 6-field 2-fluid model} \label{sec_model}

\begin{table}
\begin{ruledtabular}
\begin{tabular}{|c|}
% ine
$n = n_0 n_N$\\ % ine
% \rowcolor{LightGray}
$B = B_0 B_N$\\ % ine
$u_{\parallel} = v_A u_{\parallel \, N}$\\ % ine
% \rowcolor{LightGray}
$\nabla_{\perp} = L_{\perp}^{-1} \nabla_{\perp \, N}$ \\ % ine
$\nabla_{\parallel} X_N = L_{\perp}^{-1} \{ \psi_N, X_N \} - L_z^{-1}\partial_{z \, N} X_N$ \\ % ine
% \rowcolor{LightGray}
$\partial_t = \tau_A^{-1} \partial_{t \, N}$\\
$\tau_A = L_{\perp}/v_A$\\ % ine
$\psi = B_0 L_{\perp} \psi_N$ \\ % ine
% \rowcolor{LightGray}
$\phi = B_0 L_{\perp} v_A \phi_N$ \\ % ine
$\tau_i = T_{i \, 0} / T_{e \, 0}$ \\ % ine
% \rowcolor{LightGray}
$\Omega_i = eB_0/m_i$\\ % ine
$\rho^2_* = \frac{T_0}{m_i} \frac{1}{\Omega^2_i}  \frac{1}{L^2_{\perp}} \frac{1}{1+\tau_i} = \frac{T_{0 \, e}}{m_i} \frac{1}{\Omega^2_i}  \frac{1}{L^2_{\perp}} = \frac{\beta_{e\,0}}{2n_0}$ \\ % ine
% \rowcolor{LightGray}
$p_e = \frac{p_0}{(1 + \tau_i) \rho^2_* \Omega_i \tau_A } \, p_{e \, N} $ \\ % ine
$p_i = \frac{p_0 \, \tau_i}{(1 + \tau_i)\rho^2_* \Omega_i \tau_A} \, p_{i \, N} $ \\
% ine
% \rowcolor{LightGray}
$\bm{\mathcal{K}_3} = \bm{\mathcal{K}_2} = -(5/3) \rho^2_* \bm{\mathcal{K}_1}$\\
% ine
\end{tabular}
\caption{\label{table_normalization} Normalization used for the equations of the model. $n_0$, $B_0$, $v_A = \frac{c B_0}{\sqrt{m n_0}}$, $\Omega_i$, $\rho^2_*$ and $p_0 = p_{e \, 0} + p_{i \, 0}$ are, respectively, the values of density, magnetic field, Alfvén velocity, ion gyrofrequency, (square of) ion sound gyroradius and total pressure at the resonant surface. $L_{\perp}$ is a characteristic perpendicular length of the system, while the $\partial_z$ term only appears in 3D simulations (i.e. not in this work).}
\end{ruledtabular}
\end{table}

The 6-field fluid model used for this study was developed starting from the usual Braginskii fluid equations \cite{braginskii1965transport}. The model is capable of describing both small-scale phenomena, like interchange instabilities and the deriving turbulence, and large-scale ones, like magnetic islands and zonal fields, on time scales sufficiently long to observe the long-term development of non-linear dynamics and even phenomena arising on resistive time-scales.\\
The magnetic field is used as reference to define a parallel direction and a perpendicular direction in the system. The model being a reduced two-fluid model means that there is direction in which the fluctuations of the magnetic field are neglected, thus {$\bm{\tilde{B}} = \nabla \times (\tilde{\psi} \bm{\hat{z}}) $}, and compute the parallel derivative of a generic function $f$ keeping into account {also} the fluctuations of $\bm{\tilde{B}}$ using a Poisson bracket: $\bm{\tilde{B}} \cdot \nabla f = \{\tilde{\psi}, f \}$, defined below {(Using the total field $\psi$ in the Poisson brackets means that both equilibrium and fluctuations contribute to the term)}.\\
Most studies done in the past employed models with fewer dynamic fields, that allowed to study interchange-like turbulence in a variety of conditions \citep{yagi2007nonlinear, ishizawa2010turbulence, muraglia2011generation, agullo2017nonlinearI, dubuit2021turbislands}. The extension to a larger number of fields \cite{scott2001low1, scott2001low2, giacomin2021gbs, zholobenko2021electric, ZHU201846} can provide significant insight, especially at lower $\beta$ regimes, and in particular with what concerns the transport properties of different quantities in a given system. The model used in the present article is a reduced electromagnetic fluid model that allows to study the individual dynamics of electrons and ions by having an evolution equation for the pressure of each species, while maintaining quasi-neutrality with a single equation for the density. It also introduces the parallel velocity of the ions as a dynamic field, which is fundamental for the study of parallel dynamics (e.g. compressional sound waves).\\
The full expression of the flux terms is retained in the model, meaning that, in reduced notation, the parallel compression terms are given by the product of 3 fluctuating fields, e.g.
\begin{equation} \label{eq_pressure_parallel_compression}
p  \nabla_{\parallel} u_{\parallel} = p  \lbrace \psi , u_{\parallel} \rbrace
\end{equation}
where $\psi$ is the poloidal magnetic flux, $u_{\parallel}$ the parallel velocity, $p$ the pressure and the Poisson bracket is defined as $\lbrace f, g \rbrace = (\nabla f \times \nabla g ) \cdot b_{z}$, which in slab geometry is $ \lbrace f, g \rbrace = ( \partial_x f \, \partial_y g - \partial_x g \, \partial_y f )$, with $b_z$ being the normalized magnetic field in the $z$ direction. The presence of these non-linearities (called ``cubic terms'') can significantly affect the dynamics of the system, as shown in \cite{villa2022localized} and already introduced in \cite{villa2025zonal}, and as will be further demonstrated here.\\
Also kept in the equations is the compression of (but not the advection by) the polarization velocity ($\bm{u}_{pol} = \Omega^{-1}_i \bm{b} \times \frac{d}{dt} (\bm{u}_E + \bm{u}_{pi})$, with $\bm{b} = \bm{B}/B$, while $\bm{u}_E$ and $\bm{u}_{pi}$ are, respectively, the $E \times B$ drift and ion diamagnetic drift velocities) in the equation for the ion pressure as well as for the perpendicular dynamics, while the advection by the parallel velocity for both ions and electrons is always considered. The normalization is detailed in table \ref{table_normalization}, while further details on the derivation of the equations can be found in \cite{villa2022localized}. { A comparison of the equations used here to similar ones in other models} \cite{giacomin2021gbs, zholobenko2021electric}{ leads to believe that the phenomena observed here should be at play also in those codes, since they also keep cubic terms in the equations and the whole 6-field system is solved, though other models are usually applied to more realistic geometries and used to perform comparisons with experiments. {This would offer an interesting opportunity for future explorations of what is described in this paper and how it relates to what is presented in studies carried out with GRILLIX or GBS}\cite{zholobenko2024tokamak, giacomin2022turbulent} {. For example in one study carried out with GRILLIX}\cite{zholobenko2024tokamak}{ the authors find that their simulations lead to the appearance of KBMs near the separatrix, and these are then observed to drive the electromagnetic heatflux. The question spontaneously arises whether in these simulations one could also observe the presence of small-scale islands in the region of unstable KBM, and possibly whether they play a role also in this ``realistic'' scenario. However, considering the more involved experimental modeling of the discharges carried out in ref.}\cite{zholobenko2024tokamak}{ (inclusion of neoclassical effects, sourcing in both particles and energy, global domain, free-streaming parallel transport etc...), and, fundamentally, the difference in the parameters, even when considered locally (e.g. the normalized shear in the region where the KBM is detected in the paper is reported as $\hat{s} = r\frac{\partial_r q}{q} = 15.2$ and the value of $\beta = 0.07 \%$) one should really understand how these coalescing islands behave in a simple system before approaching such a case, which is the role of the present paper. Similar considerations apply to results presented with GBS and that has been used to look into the role played by (resistive) ballooning modes on plasma confinement in realistic geometries}\cite{giacomin2022turbulent}. {The main difference between this model and those used in the references is the fact that here the pressures are evolved instead of the temperatures, and that the vorticity equation employs the Boussinesq approximation}\cite{ross2018effect}{ to allow the determination of the electrostatic potential $\phi$ as a matrix operation rather than using an iterative solver.}\\

\begin{widetext}
\begin{eqnarray} \label{eq_psiev_def}
\partial_t \psi &=& \{ \psi, \phi \} - \frac{\Omega_i \tau_A \rho^2_*}{n} \{ \psi, p_e \} + \frac{\eta}{n} \tilde{J_{\|}}
\end{eqnarray}
\begin{eqnarray} \label{eq_omegaev_def}
 \partial_t \mathcal{W} &=& - \{ \phi, \mathcal{W} \} - \tau_i \Omega_i \tau_A \rho^2_* \{ \nabla_{\alpha} \phi , \nabla_{\alpha} p_i \} - u_{\parallel \, i} \{ \psi , \mathcal{W} \} + \{ \psi , J_{\parallel} \} \\ \nonumber
&& + \rho^2_* (\Omega_i \tau_A)^2 \bm{\mathcal{K}_1} \! \cdot \! \nabla(\tau_i p_i + p_e) + \mu \Delta_{\perp} \tilde{\mathcal{W}}
\end{eqnarray}
\begin{eqnarray} \label{eq_piev_def}
\partial_t p_i &=& \! - \frac{5}{3} p_i \{ \psi, u_{\parallel \, i} \} - \{ \phi, p_i \} - u_{\parallel \, i} \{ \psi, p_i \} - \frac{5}{3}p_i \bm{\mathcal{K}_2} \cdot \nabla \phi + \frac{5}{3} \frac{1}{\Omega_i \tau_A} T_i \{ \psi , J_{\parallel} \} \\ \nonumber
&& - \frac{5}{3} \Omega_i \tau_A \rho^2_* \Big( \tau_i p_i \bm{\mathcal{K}_2} \! \cdot \! \nabla T_i  - T_i\bm{\mathcal{K}_2} \! \cdot \! \nabla p_e \Big) + \chi_{\perp \, i} \Delta_{\perp} \tilde{T}_i + \chi_{\parallel \, i} \{ \psi , \{ \psi , \tilde{T}_i \} \}
\end{eqnarray}
\begin{eqnarray} \label{eq_peev_def}
\partial_t p_e &=& - \frac{5}{3} p_e \{ \psi, u_{\parallel \, e} \} - \{ \phi, p_e \} - u_{\parallel \, e} \{ \psi, p_e \} - \frac{5}{3} p_e \bm{\mathcal{K}_2} \! \cdot \! \nabla \phi \\ \nonumber
&& + \frac{5}{3} \Omega_i \tau_A \rho^2_* \bm{\mathcal{K}_2} \! \cdot \! \nabla(T_e p_e) + \chi_{\perp \, e} \Delta_{\perp} \tilde{T}_e + \chi_{\parallel \, e} \{ \psi , \{ \psi , \tilde{T}_e \} \}
\end{eqnarray}
\begin{eqnarray} \label{eq_nev_def}
\partial_t n &=& - \{ \phi, n \} - n \{ \psi, u_{\parallel \, e} \} - u_{\parallel \, e} \{ \psi, n \} -  n \bm{\mathcal{K}_3} \! \cdot \! \nabla \phi + \Omega_i \tau_A \rho^2_* \: \bm{\mathcal{K}_3} \! \cdot \! \nabla p_e + D \Delta_{\perp} \tilde{n}
\end{eqnarray}
\begin{eqnarray} \label{eq_upar_def}
\partial_t u_{\parallel \, i} &=& - \{ \phi , u_{\parallel \, i} \} - \left\{ \psi, \frac{u_{\parallel \, i}^2}{2} \right\} - \frac{(\Omega_i \tau_A)^2 \rho^2_*}{n}  \{ \psi, \tau_i p_i + p_e \} + U_d \Delta_{\perp} \tilde{u}_{\parallel \, i}
\end{eqnarray}
\end{widetext}

where $\psi$ is the poloidal magnetic flux, $\phi$ the electrostatic potential, $p_e$ and $T_e$ the electron pressure and temperature, $p_i$ and $T_i$ the ion pressure and temperature, $n$ the electron density (quasi-neutrality is assumed) and $u_{\parallel \, i/e}$ the ion and electron parallel fluid velocities, with 
\begin{equation} \label{eq_ue}
u_{\parallel \, e} = u_{\parallel \, i} - \frac{J_{\parallel}}{n \, \Omega_i \tau_A}
\end{equation}
The definition of the generalized vorticity is $\mathcal{W} = \Delta_{\perp} (\phi + \tau_i \Omega_i \tau_A \rho^2_* p_i)$, meaning that here the Boussinesq approximation is applied, and the parallel current density is $J_{\parallel} = \Delta_{\perp} \psi$. A ``$\sim$'' above the symbol of the field indicates its fluctuating component, e.g. $\tilde{\psi} = \psi - \psi_{eq}$. $\alpha$ in Eq. \ref{eq_omegaev_def} is an index for the sum over the perpendicular geometrical coordinates, since the explicit expression of  the second term on the RHS is (in slab geometry):
\begin{equation}
\{ \nabla_\alpha \phi , \nabla_\alpha p_i \} = \{ \partial_x \phi , \partial_x p_i \} + \{ \partial_y \phi , \partial_y p_i \}
\end{equation}
Using the definition of the generalized vorticity a condition can be imposed to have zero-vorticity equilibrium profiles, which is always satisfied in simulations unless clearly stated:
\begin{equation} \label{eq_vorticity_eq}
\partial_x \phi_{eq} = - \Omega_i \tau_A \rho^2_* \frac{\partial_x p_{i \, eq}}{n_{eq}}
\end{equation}
The curvature is defined as:
\begin{equation}
\bm{\mathcal{K}} = \left( \frac{1}{B} \nabla \times \bm{b} + \nabla \left( \frac{1}{B} \right) \times \bm{b} \right) \approx -\frac{\partial_x B_{y}}{B^2_{eq}} \bm{y}
\end{equation}
{where $B_{eq} = B_{z \, eq} + B_{y \, eq} (x)$ is the profile of the equilibrium magnetic field.}\\
The distinction of the parameters $\bm{\mathcal{K}_1}$, $\bm{\mathcal{K}_2}$ and $\bm{\mathcal{K}_3}$ is done to have more freedom in manipulating the linear spectrum, but this is irrelevant for the simulations in this work, as they are kept to their consistent values (see Tab. \ref{table_normalization}).\\
This model is implemented in the fluid code AMON \citep{poye2012dynamique}.\\
Note that keeping the cubic terms leads to the presence of additional terms, both quadratic and cubic, with respect to other models. These are the last two ``\textit{Non-Linear}'' rows in Eq. \ref{eq_terms_breakdown} (the subscript ``$eq$'' indicates its equilibrium part) and their effect on the dynamics will be partially the subject of this paper.
\begin{eqnarray} \label{eq_terms_breakdown}
p\{\psi , u_{\parallel}\} \! = \! \left[ \begin{array}{c}
\! \left\{\begin{array}{l|c}
\! Lin & p_{eq} \Big( \partial_x \psi_{eq} \partial_y \tilde{u}_{\parallel} \! - \! \partial_x u_{\parallel \, eq} \partial_y \tilde{\psi} \Big)
\end{array}\right. \\
\\
\! \! \left\{\begin{array}{l|c}
  & p_{eq} \Big( \partial_x \tilde{\psi} \partial_y \tilde{u}_{\parallel} \! - \! \partial_x \tilde{u}_{\parallel} \partial_y \tilde{\psi} \Big)\\
 \! \! \! NonLin & \tilde{p} \Big( \partial_x \psi_{eq} \partial_y \tilde{u}_{\parallel} \! - \! \partial_x u_{\parallel \, eq} \partial_y \tilde{\psi} \Big)\\
  & \tilde{p} \Big( \partial_x \tilde{\psi} \partial_y \tilde{u}_{\parallel} \! - \! \partial_x \tilde{u}_{\parallel} \partial_y \tilde{\psi} \Big)
\end{array}\right.
\end{array}
 \right.
\end{eqnarray}

Lastly, a remark about terms of the kind $1/n$, that for convenience (i.e. allowing to separate the linear from the non-linear dynamics, as the former are computed in Fourier space, while the latter in ``coordinate'' space) are implemented in the code using a first order expansion:
\begin{equation}
\frac{1}{n} = \frac{1}{n_{eq} + \tilde{n}} \approx \frac{1}{n_{eq}} - \frac{\tilde{n}}{n^2_{eq}}
\end{equation}
{this is needed because the code computes first the linear part of the equations, and then, if chosen in the input file, adds the non-linear computation, meaning that one always needs a linear expression for all terms.}
%though they have yet to show major impact in the simulations, since the profiles of density are always taken to be flat.

\section{Linear analysis of the system} \label{sec_linear}

For the simulations of this work, a 2D single-helicity system with slab geometry is considered, where the only instability present is the interchange-like fluid KBM instability, driven by the coupling of the equilibrium ion pressure gradient to the curvature of the magnetic field. Since the system in question is neither kinetic nor toroidal, strictly speaking it would be improper to refer to the instabilities as Kinetic Ballooning Modes (KBMs) {and the terminology Ideal Balloning Mode should be employed}, but since the simulated plasma has high $\beta$, the instability is driven by the coupling of magnetic field curvature and pressure gradient and the system is fully electromagnetic, the interchange instability present here is used as a proxy for KBMs, and is referred to as ``fluid KBM'', as is customary in the literature {(see reference }\cite{zocco2018strongly}{ for a comparison of the gyro-kinetic and MHD properties of said instability, that match rather well when diamagnetic terms are included in the MHD model, as is the case for the fluid model used in this paper)}.\\
In a two-dimensional system, one considers only fluctuating fields with a given helicity, that can thus only be resonant at the position where the equilibrium magnetic field also has the chosen helicity. It is then the most natural choice to use quantities at the resonant position as reference values for the normalization of the quantities in the system. To limit the number of parameters at play, the only gradients in the system are those of the ion pressure and of the magnetic field, shown in Fig. \ref{fig_eq_profiles}. The magnetic field is set by a Harris-sheet profile, chosen such that it is linearly stable to the tearing mode {(the $B_{z \, eq} \gg B_{y \, eq}$ component is considered constant and used as normalization)}, which is confirmed by computing, at the resonant position $x_0 = 0$, the parameter \citep{furth1963finite, arcis2006rigorous}:
\begin{eqnarray}
B_{y \, eq} ( x ) = A_B \tanh\left( \frac{x - x_{res}}{a_B} \right)\\
\Delta' = \lim\limits_{\epsilon \rightarrow 0} \frac{\psi' (x_0 + \epsilon) - \psi' (x_0 - \epsilon)}{\psi(x_0)} = -1.9 < 0
\end{eqnarray}
The variation of $\partial_x B_{eq}$ is done by varying the width of the hyperbolic tangent ($a_B$) in the Harris profile, and the resonant position $x_{res}$ will always be $x_{res} = 0$ in the simulations presented.\\
Notice also that the resistivity $\eta = 10^{-5}$ implies a typical time-scale for resistive reconnection $\tau_\eta = L_\perp^2/\eta \approx 10^3 \; \tau_A $ for a typical $L_\perp = 0.1$, meaning that phenomena occurring on time-scales faster than a few $\tau_\eta$ are due to turbulent dynamics.\\
The a-dimensional parameters used for the simulations are indicated in Tab. \ref{tab_typical_parameters_2D}, corresponding to a plasma with $0.64 \% \leq \beta \leq 2.56 \%$ and the simulations are run on a grid with $N_x = 512$ and $N_y=256$, with de-aliasing for the non-linear part (see App. \ref{sec_dealiasing}). Convergence of the dynamics for the chosen resolution was checked.\\
For an example of the linear growth rate spectrum of the simulations run in this work the reader is referred to Fig. \ref{fig_spectra}, where the blue dots (labeled ``CA'') correspond to simulation $\mathbb{S}1$ in section \ref{sec_generation}.\\
Because the instability is fluid KBM, if the system were perfectly symmetric (see section \ref{sec_linear_cubic}) the parities of the fields in the linear system (indicating even parity with a ``$+$" and odd parity with a ``$-$'') can be shown \citep{ishizawa2010turbulence} to be:
\begin{equation}
\Big( \psi^-, \phi^+, p_i^+, p_e^+, n^+, u_{\parallel \, i}^- \Big)
\end{equation}
In this work, the change in parity of the modes of $\psi$, which occurs during the non-linear phase, is used as a criterium to identify the formation of TDMIs, since the parity of $\psi$ in a magnetic island is even, but for interchange instabilities this field starts out as odd. Rather than using the usual definition of even and odd, based on the values of the function on either side of a reference point, here the definition takes advantage of the fact that performing a Fourier transform on the fields (i.e. for a field $f$ $f(x,y) = \sum_m f_m(x) \exp(i k_m y +\varphi(x))$) gives complex valued functions. Thus one can consider the argument of the complex Fourier mode over the radial coordinate and, if the function is even (odd) the value of the argument will be constant (changing by $\pi$), i.e. $\Delta \varphi = | \varphi (x+x_0) - \varphi (x-x_0) | = 0 \; (\pi)$ across the resonant position $x_0$. This definition has the advantage of being extensible to 3D and of making it easier to account for shifts in the resonant position due to the action of the zonal current, as will be shown. Furthermore, as detailed shortly below (see Sec. \ref{sec_linear_cubic}) the presence of the cubic terms and the definition of the temperature fluctuations breaks the exact parity one would expect from a system with symmetric gradients as the one considered here (see Fig. \ref{fig_eq_profiles}).

\begin{table}
\begin{ruledtabular}
\begin{tabular}{|ccc|cc|}
% ine
$\eta$ & $1 \cdot 10^{-5}$ &\qquad \qquad & $\rho_*^2$ & $1.66 \cdot 10^{-4}$ \\% ine
% \rowcolor{LightGray}
$\mu$ & $5 \cdot 10^{-4}$ &\qquad \qquad & $\mathcal{K}_1$ & $-0.5 \; [-1]$  \\% ine
$\chi_{\perp \, e/i}$ & $5 \cdot 10^{-5}$ &\qquad \qquad & $\mathcal{K}_2 = \mathcal{K}_3$ & $-3 \cdot 10^{-1}$ \\% ine
% \rowcolor{LightGray}
$\chi_{\parallel \, e}$ & $5.0\cdot10^{2}$ &\qquad \qquad & $L_y$ & $6 \pi$ \\% ine
$\chi_{\parallel \, i}$ & $8.0 \cdot 10^{0}$ &\qquad \qquad & $\tau_i = T_i/T_e$ & $1.0$ \\% ine
% \rowcolor{LightGray}
$D$ & $1 \cdot 10^{-4}$ & \qquad \qquad & $\Omega_i \tau_A$ & $3.12 \; [6.24]$ \\% ine
$U_d$ & $5 \cdot 10^{-5}$ & \qquad \qquad & $L_x$ & $4$\\% ine
\end{tabular}
\caption{\label{tab_typical_parameters_2D} Dissipative and dimensionless parameters used for the 2D non-linear simulations in slab geometry. The value of $\Omega_i \tau_A$ in brackets is for the higher-$\beta$ case. The value of $\mathcal{K}_1$ in brackets is for the stronger drive case.}
\end{ruledtabular}
\end{table}

\begin{figure}
\centering
\includegraphics[width=\DefFigWidth, trim=0 0 0 3cm, clip]{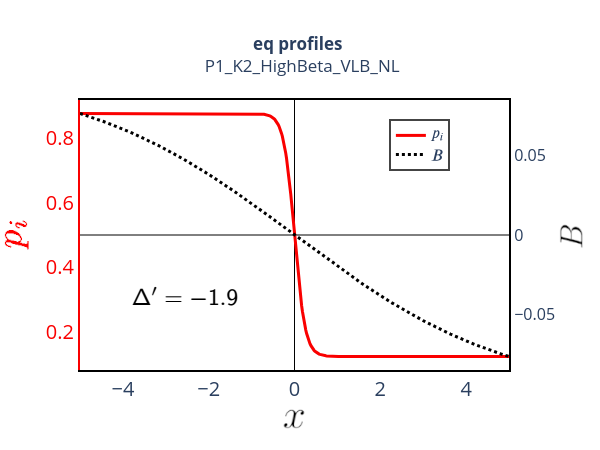}
\caption{\label{fig_eq_profiles}Equilibrium profiles for the 2D slab simulations. Profiles not shown have null gradient (except $\partial_x \phi_{eq} = -\Omega_i \tau_A \rho^2_* \partial_x p_{i \, eq}$).}
\end{figure}

{Since the most detailed analysis in this paper will be carried out for simulations $\mathbb{S}4$ and $\mathbb{S}5$, both of which have $\beta = 0.64 \%$, Fig.} \ref{fig_beta-comp} {shows the spectra of the modes around the most unstable one for simulations run with $\beta = 2.56 \%$ ($\mathbb{S}1$) and $\beta = 0.64 \%$ ($\mathbb{S}2$). As visible, aside from the slight shift in the mode number of the most unstable mode and the growth rates, expected given the significant difference in $\beta$, the spectra look remarkably similar, and the dynamics are also rather similar, except that the larger $\beta$ cases, due to their larger growth rates, reach the condition where the islands touch the radial boundaries of the simulation domain much earlier, and thus present less data to carry out the analysis.}\\

\begin{figure}[H]
\centering
\includegraphics[width=\DefFigWidth, trim=0 0 0 3cm, clip]{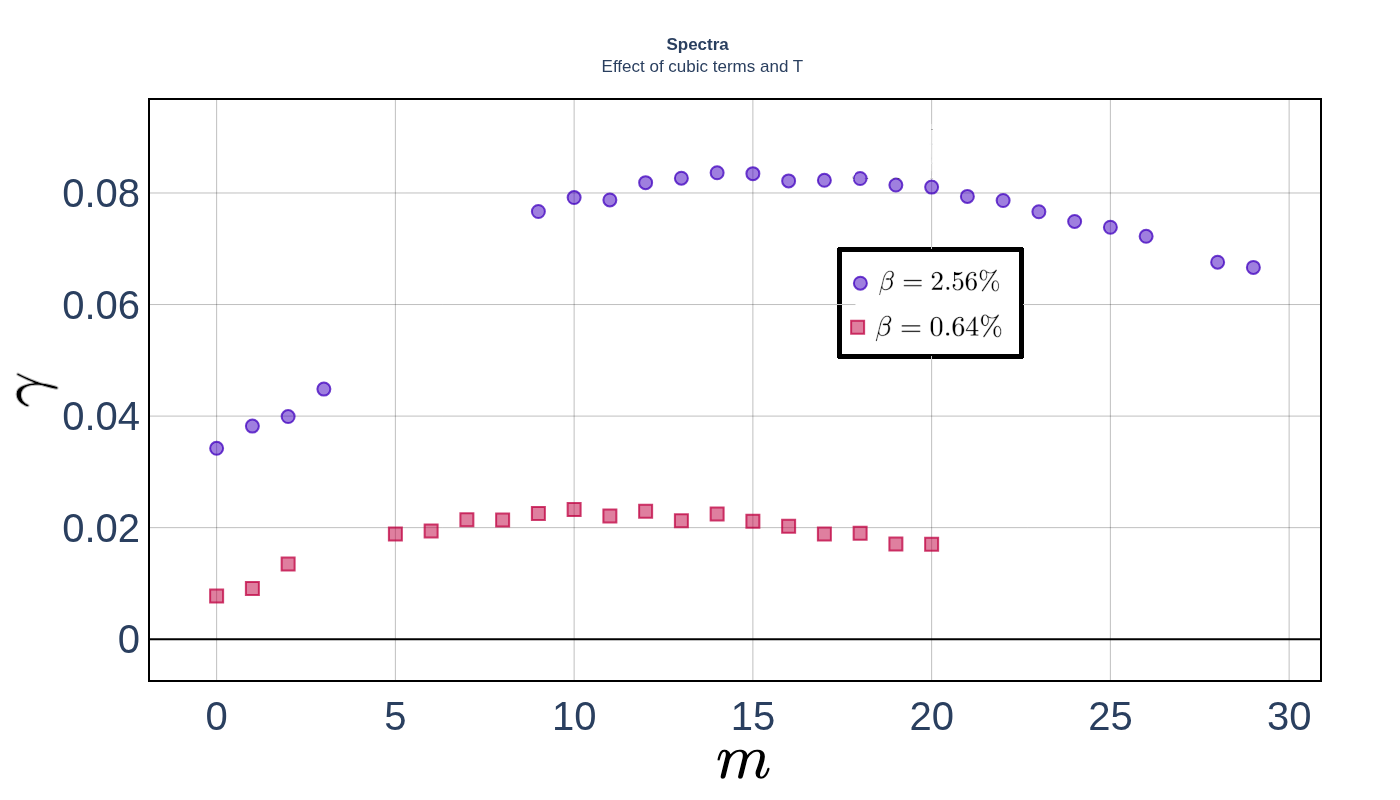}
\caption{\label{fig_beta-comp}Comparison of the growth rates of the unstable modes in simulations $\mathbb{S}1$ and $\mathbb{S}2$. Only shown the part of the spectrum around the most unstable modes. Missing points are due to the results coming from non-linear simulations, where the growth rates for modes are not always clearly identifiable.}
\end{figure}

\subsection{Sources of asymmetry in the system} \label{sec_linear_cubic}
Further attention needs to be dedicated to the role played by the cubic terms and the equilibrium profiles in general in the system, as they act as a ``non-explicit'' source of asymmetry. Addressing this in the single-helicity linear system allows to approach the problem with more clarity.\\
One will notice how in Fig. \ref{fig_eq_profiles}, the gradients of the profiles are symmetric about the resonance. This, in a system that only retains quadratic terms, would imply that the dynamics of the perturbed fields will have a certain symmetry around the resonance. This is true in slab geometry and would be untrue, for example, if one were to move to cylindrical geometry, where the radial dependence of the poloidal derivative introduces asymmetries in the system  ``implicitly'', even if all gradients were symmetric (see e.g. \cite{poye2011asymmetry, poye2013global, smolyakov2013higher}).\\
In the present model, even in slab geometry and for symmetric equilibrium gradients, there are 2 ``implicit'' sources of asymmetry: the cubic terms and the temperature fluctuations.\\
The first source is related to the fact that the cubic terms involve the full profile of the background fields in the linearized Poisson brackets, e.g. from Eq. \ref{eq_piev_def}, for a given mode $m$
\begin{eqnarray}
\partial_t p_{i \, m} & = & ... + p_i \{ \psi, u_{\parallel \, i} \} \\
\nonumber \partial_t p_{i \, m} & = & ... + p_{i \, eq} (x) \Big( \partial_x \psi_{eq} (x) \partial_y u_{\parallel \, i \, m} (x) \\
& & - \partial_x u_{\parallel \, i \, eq} (x) \partial_y \psi_m (x) \Big)
\end{eqnarray}
Thus, cubic terms determine in the system a high-pressure region and a low-pressure region, that enter explicitly in the equations through the consideration of the full profile of the pressure rather than just the reference or normalization value.\\
The second ``implicit'' source of asymmetry is the definition of the temperature fluctuations, that can follow two alternative paths, independently of whether or not one retains cubic terms in the model. The first possibility is to take into account the equilibrium profiles of pressure and density in the computation of the temperature fluctuations (the subscript ``s'' indicates the plasma species ``i'' and ``e''):
\begin{equation} \label{eq_T_asymm}
\tilde{T}_s = \tilde{p}_s \frac{1}{n_{eq}(x)} - \frac{T_{eq \, s}(x)}{n_{eq}(x)} \tilde{n}
\end{equation}
meaning that the temperature fluctuations have larger amplitude on the high-pressure side, and smaller amplitude on the low-pressure side, but, crucially, always average to $0$ (in statistical terms, their variance changes but their average doesn't over the radial coordinate). This is essentially a way to consider the global properties of the system rather than expanding around a local resonance.\\
The second possibility is to define the temperature fluctuations in such a way that they are uniform in amplitude over the whole radius, as is the case for all the other fields, that are initialized with random initial conditions:
\begin{equation} \label{eq_T_symm}
\tilde{T}_s = \tilde{p}_s \frac{1}{n_0} - \frac{T_{0 \, s}}{n_0} \tilde{n}
\end{equation}
where values with sub-script $0$ are single reference values, not full profiles. One can think of this latter case, Eq. \ref{eq_T_symm}, as the local limit of the global case in Eq. \ref{eq_T_asymm}.\\
The effect on the linear system of these two elements is shown in Fig. \ref{fig_asy_all}, that summarizes the results for some variations of the same linear simulation, corresponding to simulation $\mathbb{S}1$ in section \ref{sec_generation}. The labels in these figures are such that results marked by labels that contain the letter ``A" (for ``asymmetric'') use the definition of the temperature as per Eq. \ref{eq_T_asymm}, whereas labels that contain the letter ``S" (for ``symmetric'') use the definition in Eq. \ref{eq_T_symm}. Labels containing letter ``C" (for ``cubic") are simulations run with cubic terms, thus the full profile of the equilibrium quantities plays a role in the linear dynamics, while labels containing letter ``Q" (for ``quadratic") only keep the reference value of the profile in question.\\
Fig. \ref{fig_spectra} shows the impact on the linear spectrum of the elements mentioned. Any added degree of asymmetry renders the system more unstable, but doesn't alter the range of modes at which the instabilities develop. The wave-number of the most unstable mode is around $m^* \approx 10$ and is only slightly affected by these variations.\\
Fig. \ref{fig_asymmetry} shows how the eigenmodes of $\psi_{10}$ (mode chosen because very close to $m^*$ in all cases) are impacted by the sources of asymmetry. Notice in particular how adding asymmteries increases the relative magnitude of the right-hand (low-pressure) side of the eigenmode. This will be impactful in the non-linear dynamics (see section \ref{sec_dynamics}), and is notably an effect that opposes the effects of cylindrical geometry: cylindrical geometry tends to render the eigenmodes of the fields larger inwards (on the high-pressure side), due to the $1/r$ dependence of the poloidal derivatives.

\begin{figure}
\begin{subfigure}{\DefFigWidth}
\centering
\includegraphics[width=\textwidth, trim=0 0 7em 3cm, clip]{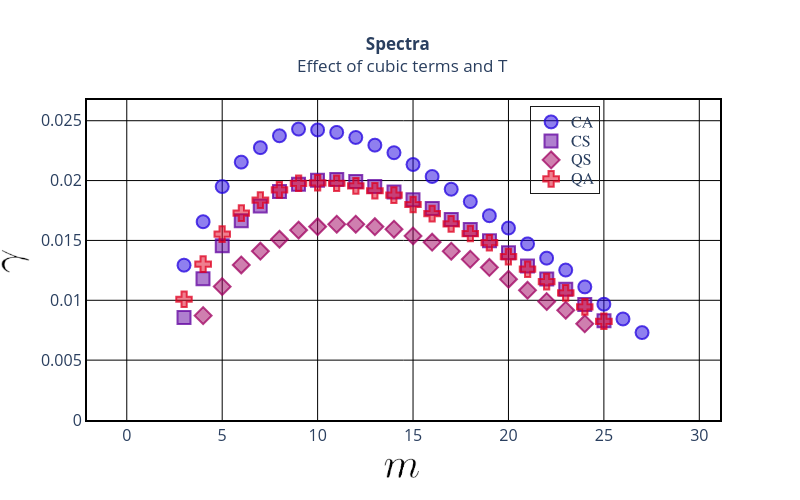}
\caption{\label{fig_spectra} Unstable portion of the spectrum in linear simulations with different contributions from the cubic terms and different definitions for the temperature fluctuations.}
\end{subfigure} \\
\begin{subfigure}{\DefFigWidth}
\centering
\includegraphics[width=\textwidth, trim=0 0 7em 3cm, clip]{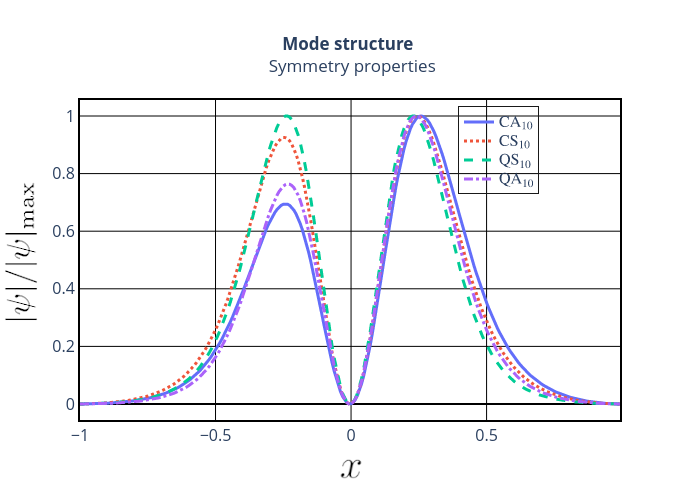}
\caption{\label{fig_asymmetry} Effect of cubic terms and temperature profile on the eigenfunctions of mode $\psi_{m=10}$. Note that the eigenmode is odd, but plotting its absolute value gives even-looking functions.}
\end{subfigure}
\caption{\label{fig_asy_all} Effect of the implicit sources of asymmetry on the linear properties of the system. Labels ``C'' (for ``cubic'') and ``Q'' (for ``quadratic'') refer respectively to simulations run with and without cubic terms. Labels ``S'' (for ``symmetric'') and ``A'' (for ``asymmetric'') refer to the definition of the temperature fluctuations used, with ``S'' referring to Eq. \ref{eq_T_symm}, and ``A'' to Eq. \ref{eq_T_asymm}.}

\end{figure}

Throughout this work, unless clearly stated, both cubic terms and ``asymmetric'' temperature fluctuations (Eq. \ref{eq_T_asymm}) are considered.%, though insight into the effect of removing cubic terms are included (see discussion in sections \ref{sec_generation_cubic} and \ref{sec_dynamics_cubic}).

\section{Generation of magnetic islands by KBM-like turbulence} \label{sec_generation}

Non-linear simulations are run taking the parameters indicated in Tab. \ref{tab_typical_parameters_2D} and scanning the parameter space as indicated in Tab. \ref{tab_sims_slab}. For the most unstable simulations ($\mathbb{S}1$ and $\mathbb{S}6$ in particular) simulations with identical parameters but larger radial domain ($L_x = 10$) were also run because the TDMIs quickly reached the boundaries of the domain. While this did not solve the issue of the island reaching the boundary, it allowed the simulation to run for longer times before results became invalid. When simulations with larger radial boxes are considered this is stated.\\
The focus is on simulations $\mathbb{S}4$ and $\mathbb{S}5$, since they are very similar overall, but $\mathbb{S}5$ shows the formation of magnetic islands, while $\mathbb{S}4$ doesn't. All simulations that show the formation of magnetic islands have similar dynamics, and once the time-scales are normalized the dynamics are seen to also happen on similar timescales. In Fig. \ref{fig_dominant_over_time}, for different simulations, the most energetic magnetic mode over time is plotted, where the time is normalized to the characteristic time-scale $\tau^* = 1/ \gamma^*$ identified by the most unstable mode. It is stressed that the linearly most unstable mode does not necessarily correspond to the most energetic mode in the non-linear phase. Notice how all simulations in Fig. \ref{fig_dominant_over_time} have a similar duration for the initial non-linear phase where the linearly most unstable mode is dominant, and the transition to the large scale modes happens for all simulations at a similar time and with similar duration in normalized units. Interestingly $\mathbb{S}4$ settles, until $t \cdot \gamma^* \approx 125$, at a mode-number roughly double that of the other simulations, possibly indicating that the presence of TDMIs facilitates the inverse cascade.\\
For simulations where magnetic islands formed, this was observed to happen through a progressively slow coalescence process. In Fig. \ref{fig_dominant_over_time} this is visible as the sharp descending mode-number phase all simulations have at $t \approx 50 \tau^*$, while later changes in energetically dominant mode-number $m_{\max}$ become much slower. This is where most of the coalescence process takes place, as small-scale islands merge together to form progressively larger scale ones. Looking in detail at the isocontours of $\psi$, in the early non-linear phase (see Fig. \ref{fig_koalabuddies}) the linear mode structures are expanded radially, while initially maintaining the parities determined by the linear properties of the interchange instability. How much these structures expand seems to be the crucial element in determining whether or not magnetic islands are going to form. Early in the non-linear phase, in simulations where the coalescence takes place, the structures at the unstable scales change their configuration forming small magnetic-island-like structures (visible in Fig. \ref{fig_koala_late}, though in this figure they are already at a lower poloidal wave-number than the one of the linear instability), meaning that there is a change in parity in the modes. At the same time reconnection sites, along with systems of nested flux-surfaces become visible. These small-scale structures begin to undergo a coalescence process, merging into larger and larger structures, in what is essentially the ``real-space'' countepart of the usual inverse energy cascade of Fourier modes in single-helicity systems. As shown in Fig. \ref{fig_NL_spectra_nodiff}, the time-scale of the non-linear inverse cascade and the re-partition of the energy are rather similar in the two cases ($\mathbb{S}4$ and $\mathbb{S}5$), as one would expect from the linear setup and the properties of the instability being so similar (see Tab. \ref{tab_sims_slab}). Fig. \ref{fig_NL_spectra_nodiff_detail} highlights the fact that the mechanism is an inverse cascade rather than non-local mode coupling (as was the case in other work \cite{yagi2007nonlinear, muraglia2011generation, agullo2017nonlinearII, dubuit2021turbislands}) because the spectra at $t=10000 \tau_A$ follow a power law rather than showing isolated peaks, that are never observed, not even in the late linear phase, where direct mode beating should be more active. This power law has different exponents for the simulation that forms TDMIs, $\mathbb{S}5$ (exponent $-2.44 \approx -12/5$), and for the simulation that doesn't form TDMIs, $\mathbb{S}4$ (exponent $-1.77$). The latter can be compared to the value $-1.66 = -5/3$ of Kolmogorov-like turbulence \citep{kida1987kolmogorov}, expected for hydrodynamic turbulence, or for Golreich-Sridhar \citep{sridhar1994toward} {strong} Alfv\'enic turbulence in MHD for perpendicular wavenumbers. Given the nature of the system at hand (relatively high-$\beta$, single-helicity, MHD), the Goldreich-Sridhar picture is probably the one that fits best. This is rather consistent with the idea that in simulations where coalescence is not occurring the non-linear dynamics at play are more in line with those already explored in the literature, and the non-linearities appear on top of a background that is affected by the fluctuations{, but is still the dominant component of the system (i.e. the parallel and perpendicular directions are not excessively disturbed by the dynamics)}. On the other hand, the index $-12/5$ {comes close to the scaling for the ``tearing-mediated regime'' ($-11/5$)}\cite{cerri2022turbulent} {, which would be consistent with the critical role played in these simulations by the TDMIs, though in these simulations there is no transition from the inertial to the tearing-mediated regime due to the dynamic range, so comparisons can be qualitative at best. Still, a fundamental novelty remains, and it is} the presence of cubic terms. {This is also why comparison with other work on coalescing magnetic islands} \cite{zhou2021statistical}{ is not straight forward, (aside from the difference in the initial setup), though future investigations might reveal deeper connection between these works. Indeed, also further comparison to the tearing-mediated turbulent regime would constitute a natural extension of this work, seeing how the cubic terms might affect the distinction of the different regimes in the inertial range, though this extends beyond the scope of the present work, requiring, among others, much more extended dynamic ranges. The crucial point is that} what determines the presence of TDMIs might also appear as a feature of the energy dynamics, but at a fundamental level is due to a change in the structures of the modes of $\psi$ to have dominant even parity, as one would expect in the presence of a magnetic island. If the system allows the modes to change parity the formation of TDMIs through coalescence becomes possible.\\
This latter point is shown in Fig. \ref{fig_NL_eigenmode}, where the radial structure of the mode $\psi_{m = 12}$ is plotted (with focus on the region around the resonance), using the color at each radial position to represent the phase difference between the mode at that particular radial position and the phase at the resonant position. For a function with purely even parity the phase would be constant across the point $x_0=0$ (here the resonance), i.e. $\Delta \varphi = \varphi( x_0-\varepsilon ) - \varphi (x_0+\varepsilon) = 0$, which is the behaviour one would expect in the case of formation of magnetic islands by tearing modes, for example, while for a purely odd function the phase difference would be $\Delta \varphi = \pi$. Since the system is non-linear, one should not expect find these exact values, and this analysis aims at determining the dominant component of the modes. In Fig. \ref{fig_NL_eigenmode} one can notice how the lower shear simulation ($\mathbb{S}5$) has a broad region of constant phase, that corresponds to the island-like structures in Fig. \ref{fig_koala_late}. The simulation with stronger shear, on the other hand, has an inversion of the phase at $ x \approx 0.2$. The interplay between the magnetic shear and the plasma $\beta$ is the crucial element in determining the behaviour of the system when it comes to the formation of TDMIs through the mechanism described here \citep{villa2025zonal}.\\
All cases run, except $\mathbb{S}4$ and $\mathbb{S}2$, showed the formation of TDMI and at least the initial phases of coalescence. Since the coalescence process can take very long times, not all simulations were run till the formation of the largest scale modes was observed, but once the coalescence process has started it is reasonable to expect it will complete the inverse cascade given enough time. In Fig. \ref{fig_avg_parities} a comparison of the average parities over time for $\mathbb{S}4$ (Fig. \ref{fig_avg_parities_1}) and for $\mathbb{S}5$ (Fig. \ref{fig_avg_parities_2}) are shown. These averages are obtained by considering the phases as in Fig. \ref{fig_NL_eigenmode} and averaging the phase difference everywhere the amplitude of the eigenmode is $\vert \psi \vert^2 / \vert \psi \vert^2_{\max} > 0.2$ (it was verified that lowering the ratio further didn't affect the results). The average over time of the absolute value for each mode is then written explicitly in the last column in Fig. \ref{fig_avg_parities}, and one can notice how, consistently, the average phase differences for the simulation that shows the formation of TDMIs, $\mathbb{S}5$, are lower at the largest scales. It must be stressed that these changes in parity occur at the same mode numbers as the linear instabilities. This method of quantifying the phases of the modes has the advantage of not relying on the resonant surface remaining fixed throughout the simulations, which is not the case due to the effect of the zonal current.\\
In the most unstable simulations the growth of the magnetic island is unbounded, and causes the separatrix to reach the boundaries of the box. Here the separatrix is defined as the flux-surface identified by isolating the most energetic mode at a given time, finding the minimum value of $\psi_{m_{\max}}$, and then tracing the isocontour of this value over the whole field, which explains why in figures the X-points are not always clearly identified.\\ Unfortunately, the time when this happens is also around the time when the transition from mode $m=2$ to mode $m=1$ happens, so that observing this transition proves challenging in simulations. The reason for this unbounded growth is unclear, especially in the cases presented here of slab geometry with current profiles stable to the tearing instability. The global profile effects described in \cite{poye2013global} are not at play here: the choice of 2D slab geometry, along with the magnetic profile chosen to be stable against the tearing mode, do not introduce additional driving flows that can drive the growth of the magnetic island far from the resonance. Thus both the generation and the growth of the magnetic island are due to the turbulence itself. Simulations with weaker instabilities did not show such unbounded growth, though in these cases the growth of the island is so slow that it wasn't possible to assess whether a saturation mechanism comes into play at later times.\\
Notice also that $\mathbb{S}5$, where TDMIs form, has weaker magnetic shear than $\mathbb{S}4$, (that is further weakened by the non-linear dynamics, as shown in section \ref{sec_dynamics}, Fig. \ref{fig_zonal_current}) which is less favorable to the destabilization of the tearing mode. Trying to attribute the formation of these magnetic islands to the non-linear de-stabilization of the tearing mode would contradict this observation.

\begin{figure}
\centering
\includegraphics[width=\DefFigWidth, trim=0 0 7em 3cm, clip]{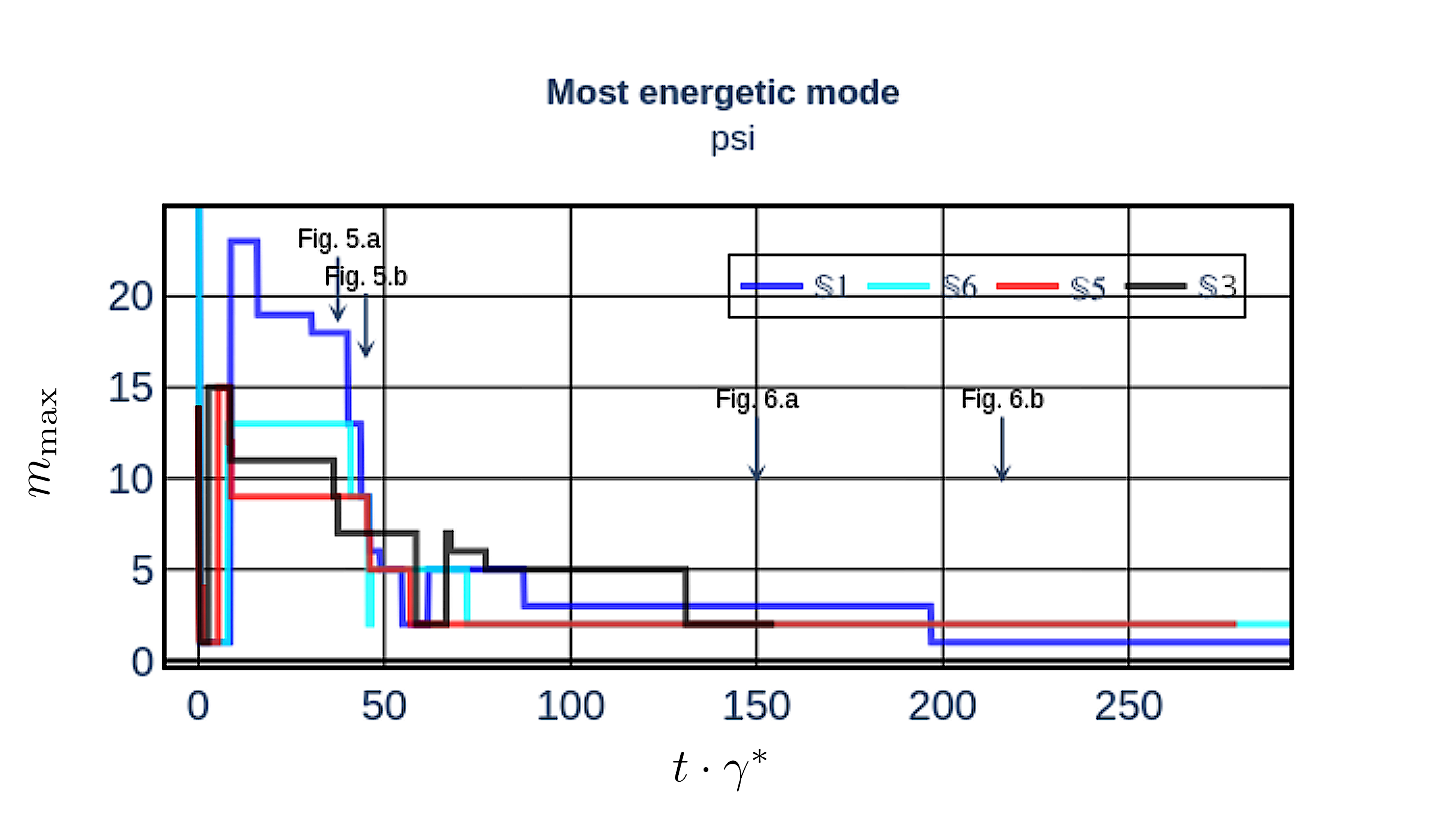}
\caption{\label{fig_dominant_over_time} Dominant mode over time of the magnetic energy ($\int_{L_x} \psi_m \cdot J_{\parallel \, m} dx$) for different simulations across the parameter range in Tab. \ref{tab_sims_slab}. The time is normalized to the linear time-scale of each simulation individually ($\tau^* = 1/\gamma^*$).}
\end{figure}

\begin{table}
\begin{ruledtabular}
\begin{tabular}{|c|c|c|c|c|c|c|}
% ine
\textbf{Name} & $\bm{\beta} \: (\%)$ & $\bm{\mathcal{K}_1}$ & $\bm{\partial_x B_{eq}}$ & $\bm{\gamma}^* \: (1/\tau_A)$ & $\bm{m}^*$ & \textbf{TDMI present} \\
$ \mathbb{S}1 $ & $2.56$ & $-1$ & $0.02$ & $0.083$ & $15$ & yes\\% ine
% \rowcolor{LightGray}
$ \mathbb{S}2 $ & $0.64$ & $-1$ & $0.02$ & $0.024$ & $11$ & no\\% ine
$ \mathbb{S}3 $ & $2.56$ & $-0.5$ & $0.02$ & $0.049$ & $13$ & yes\\% ine
% \rowcolor{LightGray}
$ \mathbb{S}4 $ & $0.64$ & $-0.5$ & $0.02$ & $0.015$ & $12$ & no\\% ine
$ \mathbb{S}5 $ & $0.64$ & $-0.5$ & $0.01$ & $0.018$ & $12$ & yes\\% ine
% \rowcolor{LightGray}
$ \mathbb{S}6 $ & $0.64$ & $-1$ & $0.01$ & $0.033$ & $13$ & yes\\% ine
\end{tabular}
\caption{\label{tab_sims_slab}Summary of parameters varied in the non-linear slab simulations with respect to what indicated in section \ref{sec_linear}.}
\end{ruledtabular}
\end{table}

\begin{figure}
\centering
\begin{subfigure}{.45\textwidth}
\includegraphics[width=\textwidth, trim=0 0 7em 3cm, clip]{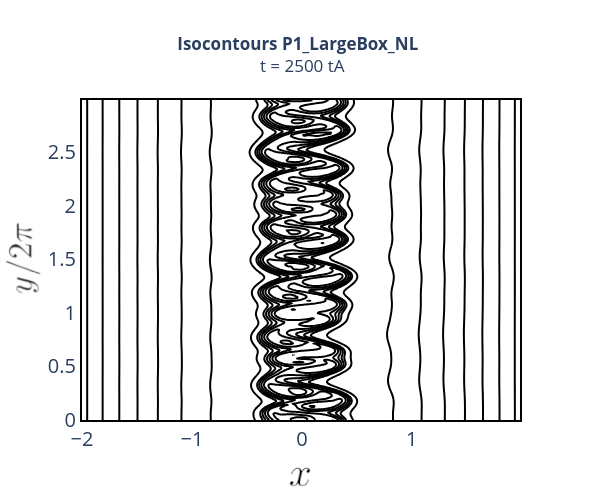}
\caption{\label{fig_nokoala}$\mathbb{S}4$: $\gamma^* = 0.015$, $\partial_x B_{eq} = 0.02$}
\end{subfigure} \, 
\begin{subfigure}{.45\textwidth}
\includegraphics[width=\textwidth, trim=0 0 7em 3cm, clip]{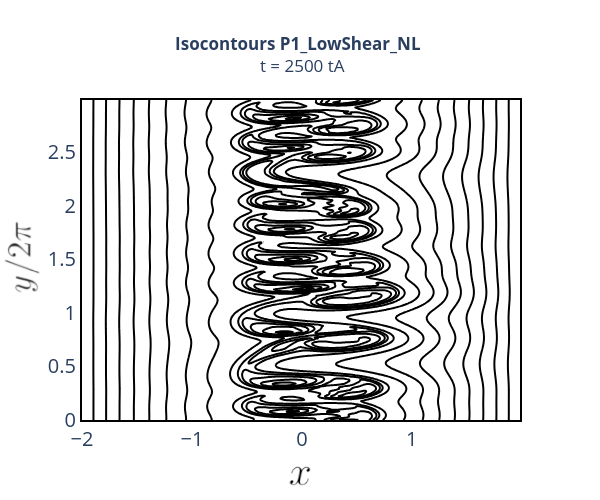}
\caption{\label{fig_koala}$\mathbb{S}5$: $\gamma^* = 0.018$, $\partial_x B_{eq} = 0.01$}
\end{subfigure}
\caption{\label{fig_koalabuddies} Isocontours of $\psi$ at $t=2500 \; \tau_A$ taken from fluid simulations run with AMON where both cases have similar linear spectra but different magnetic shears. Early non-linear phase showing turbulent structures with mostly odd parity in both simulations.}
\end{figure}

Notice the asymmetry of the structures both in Fig. \ref{fig_koala_late} and in Fig. \ref{fig_NL_eigenmode} and how it can ben related to the asymmetry highlighted in the discussion of Fig. \ref{fig_asymmetry}. %Indeed a simulations run with the same parameters as $\mathbb{S}1$ but without cubic terms gives structures that are much more symmetric and don't lead to the formation of magnetic islands (see Fig. \ref{fig_isoc_nocubic}).
This, combined with the remarks about the width of the radial structures distinguishing $\mathbb{S}4$ and $\mathbb{S}5$, leads to think that a sufficiently high degree of asymmetry in the system is favourable to the formation of magnetic islands by turbulence, as it facilitates the transition of the mode structures from interchange-like to tearing-like. Both simulations have the asymmetric definitions for the temperature (see Eq. \ref{eq_T_asymm}), but $\mathbb{S}5$ has lower magnetic shear, leading to radially-wider modes, where the asymmetry can play a more important role. Such asymmetry can be derived from the geometry, the cubic terms, or from the non-linear dynamics of the system introducing a ``seed'' asymmetry through the fluctuations. Notice, however, that in \cite{muraglia2011generation, agullo2017nonlinearI} the system was exactly symmetric and the formation of TDMI was still observed, so the asymmetry is facilitating the formation of TDMIs, but is not necessary, as will be clarified shortly.

\begin{figure}
\centering
\begin{subfigure}{.45\textwidth}
\includegraphics[width=\textwidth, trim=0 0 7em 3cm, clip]{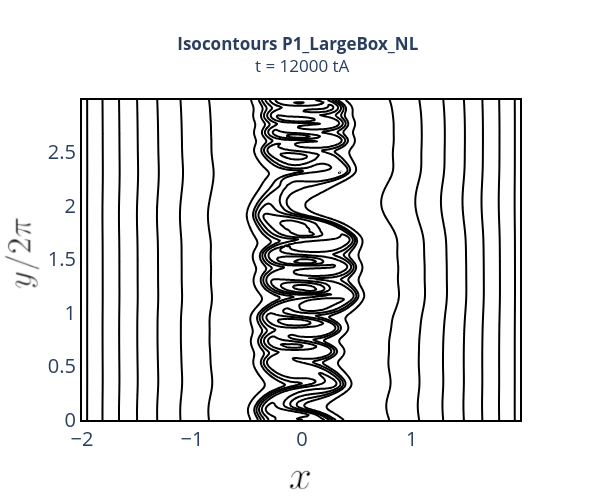}
\caption{\label{fig_nokoala_late}$\mathbb{S}4$: $\gamma^* = 0.015$, $\partial_x B_{eq} = 0.02$}
\end{subfigure} \, 
\begin{subfigure}{.45\textwidth}
\includegraphics[width=\textwidth, trim=0 0 7em 3cm, clip]{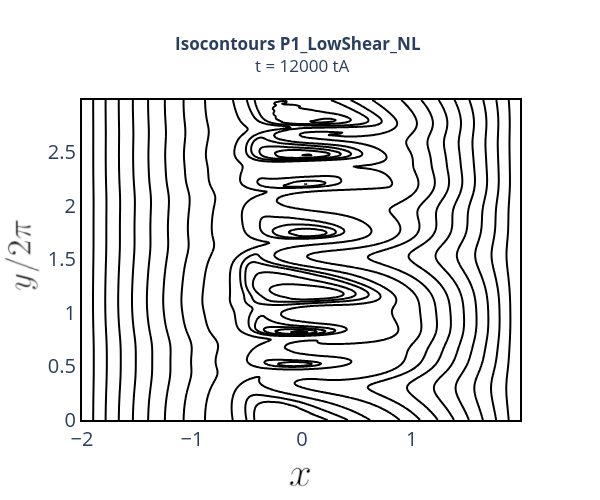}
\caption{\label{fig_koala_late}$\mathbb{S}5$: $\gamma^* = 0.018$, $\partial_x B_{eq} = 0.01$}
\end{subfigure}
\caption{\label{fig_koalabuddies_late} Isocontours of $\psi$ at $t=10000 \; \tau_A$ for $\mathbb{S}4$ and $t=12000 \; \tau_A$ for $\mathbb{S}5$ taken from fluid simulations run with AMON where both cases have similar linear spectra but different magnetic shears. Late non-linear phase showing appearance of small-scale magnetic islands in $\mathbb{S}5$ and not in $\mathbb{S}4$.}
\end{figure}

\begin{figure}
\centering
\begin{subfigure}{.48\textwidth}
\includegraphics[width=\textwidth, trim=0 0 7em 3cm, clip]{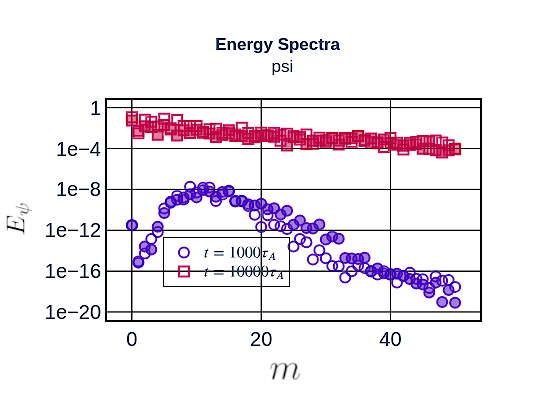}
\caption{\label{fig_NL_spectra_nodiff} Different shapes correspond to different times.}
\end{subfigure}\,
\begin{subfigure}{.48\textwidth}
\includegraphics[width=\textwidth, trim=0 0 7em 3cm, clip]{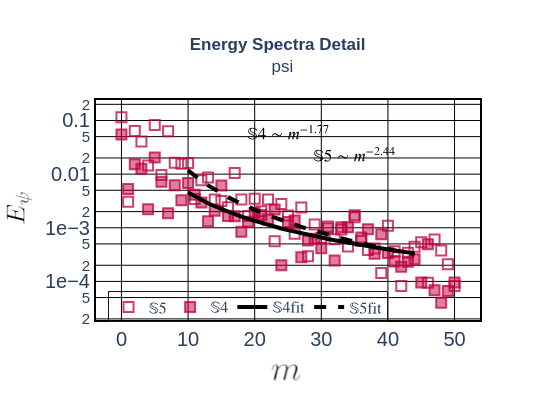}
\caption{\label{fig_NL_spectra_nodiff_detail} Spectra at $t = 10000 \tau_A$. The dashed and solid lines are obtained by fitting the data, and the corresponding power laws are indicated in the figure.}
\end{subfigure}
\caption{Comparison of the non-linear energy spectra of the field $\psi$ in the non-linear phase of simulations $\mathbb{S}4$ and $\mathbb{S}5$. Empty symbols are for the simulation that undergoes coalescence ($\mathbb{S}5$), full symbols for the one that doesn't ($\mathbb{S}4$).}
\end{figure}

\begin{figure}
\centering
\includegraphics[width=\DefFigWidth, trim=0 0 0 2.75cm, clip]{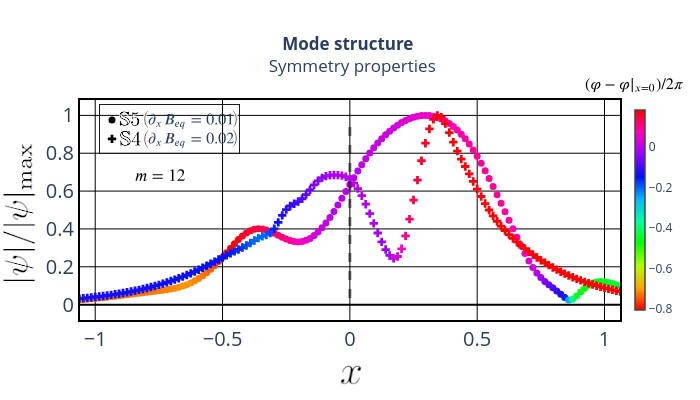}
\caption{\label{fig_NL_eigenmode}Radial structure and phases of the $\psi$ function for two different simulations with same parameters but different magnetic shear. $\mathbb{S}5$ (with $\partial_x B_{eq} = 0.01$) undergoes coalescence, $\mathbb{S}4$ (with $\partial_x B_{eq} = 0.02$) doesn't. Notice that for $\mathbb{S}5$ the real part of the function has more even-like parity, which is what one would expect for tearing modes.}
\end{figure}

\begin{figure}
\centering
\begin{subfigure}{.48\textwidth}
\includegraphics[width=\textwidth, trim=0.5cm 0 0.75cm 2.75cm, clip]{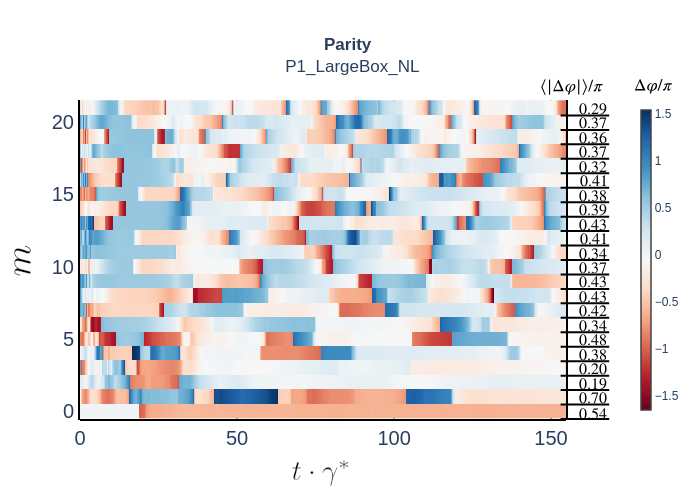}
\caption{\label{fig_avg_parities_1}$\mathbb{S}4$: $\gamma^* = 0.015$, $\partial_x B_{eq} = 0.02$}
\end{subfigure} \\
\begin{subfigure}{.48\textwidth}
\includegraphics[width=\textwidth, trim=0.5cm 0 0.75cm 2.75cm, clip]{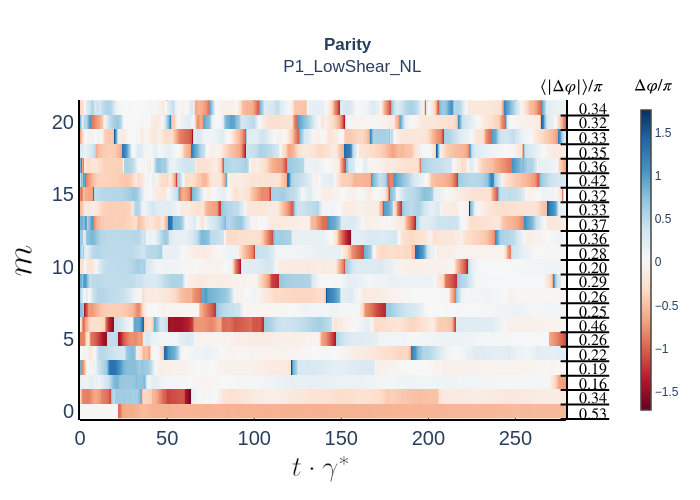}
\caption{\label{fig_avg_parities_2}$\mathbb{S}5$: $\gamma^* = 0.018$, $\partial_x B_{eq} = 0.01$}
\end{subfigure}
\caption{\label{fig_avg_parities} Comparison of radially averaged phase differences for relevant modes in simulations $\mathbb{S}4$ and $\mathbb{S}5$. The phase difference of the mode structure at each radial location with respect to the same mode at the resonant position is computed and then averaged for all radial points where $\vert \psi \vert^2 / \vert \psi \vert^2_{\max} > 0.2$. The column between the colour map and the colorscale indicates the average over time for each mode. $\mathbb{S}5$ (Fig. \ref{fig_avg_parities_2}) has consistently lower phase variation over the radius with respect to $\mathbb{S}4$ (Fig. \ref{fig_avg_parities_1}), which makes for modes that have parity closer to that expected for a tearing mode.}
\end{figure}

Here a crucial difference between the results in the present paper and those present in the literature \citep{yagi2007nonlinear, muraglia2011generation, agullo2017nonlinearII, dubuit2021turbislands} is further stressed. In former studies, the TDMIs that formed were usually the direct product of mode beating of neighbouring modes in the quasi-linear phase. Concretely, this means that for an instability most unstable at wave-number $m^*$, in the non-linear phase the coupling of this mode with the other most unstable modes, that sit at $m^* \pm 1$, generates directly the $m=1$ mode. In those systems, this mechanism is sufficient to directly form an $m=1$ magnetic island. In the simulations done for this study, this process occurs all the same, and modes with $m \leq 4$ show tearing parity from the moment non-linearities start to play a role, but these modes would not be relevant in the system if the change in parity of the smaller-scale modes and the coalescence described thus far didn't take place. These small-scale modes with changed parity undergo the inverse cascade (over the small dynamic range explored here) and deliver energy to the sub-dominant low-$m$ modes, that eventually manage to form dynamically relevant large-scale magnetic islands. This has been verified directly by running a simulation equivalent to $\mathbb{S}1$ without the cubic terms (both linearly and non-linearly), where the low-$m$ modes show tearing parity, but no island appears in any relevant way in the system, that saturates to a state similar to what has been described and shown for $\mathbb{S}4$.\\
This is where the zonal fields enter the discussion, as the change in parity of the modes with $m \approx m^*$ can be understood by focusing, as an example, on the pressure term in Ohm's law: $\frac{\Omega_i \tau_A \rho_*^2}{n} \{ \psi, p_e \}$. In an interchange unstable system the parity of $\psi_{m^*}$ is odd, and non-linear dynamics lead to the formation of modes $m=0$ of pressure and density that also have odd parity (leading to profile flattening). The Poisson bracket of two odd functions gives an odd function, but this is then multiplied by $n_{m=0}$, thus leading to an even term, at the same $m^*$ of the linear instability. This leads to the change in parity of the unstable mode.\\
Thus the element that determines the different behaviour in these simulations with respect to those in the literature is the dynamics of the zonal fields, accounted for through the cubic terms.%, that here can reach larger amplitudes than in other models. In any case, the interest lies in the fact that this is an alternative path to the formation of large-scale TDMIs, albeit a slower one.

\section{Dynamics of turbulence driven magnetic islands} \label{sec_dynamics}

The non-linear dynamics of the system are dominated by the interaction among the structures at the largest scales in the system. These structures are the TDMI, the zonal current (i.e. mode $\psi_{m=0}$) and the zonal flow (i.e. mode $\phi_{m=0}$). At first, the focus is on a simulation with the same parameters as $\mathbb{S}1$ but with $L_x=10$ in order to give the island more room to grow before interacting with the boundaries of the system.\\

\subsection{Non-linear coalescence}
As visible in Fig. \ref{fig_zonal_flow}, the poloidal flow has two regions of strong shear. The first region is localized at the resonance, where it is expected to be, both from the equilibrium condition (see Eq. \ref{eq_vorticity_eq}), and from the fact that the instability drives the zonal flow at the resonant position \citep{diamond2005zonal}. The second region is where the magnetic island has its maximum radial width, and thus moves over time as the island expands. This latter coincidence between a region of strong flow shear and maximal extension of a magnetic structure is of particular interest, as it raises the question of whether the sheared flow acts to limit the width of the island, or whether the island localizes the sheared flow where it has its maximum width.\\
Fig. \ref{fig_zonal_current} shows instead the effect of the fluctuations of $\psi$ when poloidally averaged, with the formation of a radial extension of roughly $\Delta x \approx 0.67$ where both the magnetic field $B_y$ and its shear are very weak. For this structure, the same question as for the sheared flow arises: whether the magnetic island growth is responsible for the formation of this low-shear region or if the formation of this region allows the growth of the island.\\
In order to study both of these aspects, the focus is shifted to simulation $\mathbb{S}6$, with radial width $L_x=10$, and its variants summarized in Tab. \ref{tab_variations_zonal}. Indeed, here $\mathbb{S}6$ is re-named to indicate the simulation with larger radial box for convenience. $\mathbb{S}6.a$ and $\mathbb{S}6.b$ have variations of the value of the dissipation on mode $\phi_0$, through the parameter $\mu_0$ (the other modes have the dissipative parameter $\mu$ indicated in Tab. \ref{tab_typical_parameters_2D}), i.e.
\begin{equation}
\mu \Delta_\perp \mathcal{W} = \mu_0 \Delta_\perp \mathcal{W}_0 + \mu \sum\limits_{m=1}^{N_y / 4} \Delta_\perp \mathcal{W}_m
\end{equation}
These simulations are run as entirely new simulations from the beginning. $\mathbb{S}6.c$ and $\mathbb{S}6.d$ are obtained by taking simulation $\mathbb{S}6.a$ and setting, respectively, only the fluctuations $\tilde{\phi}_0 = 0$ or both $\tilde{\phi}_0 = 0$ and $\partial_x \phi_{eq} = 0$ during the evolution from a certain time-point on (this time-point is close but different for each and in the late non-linear phase, where the mode $\psi_2$ is well developed, see Fig. \ref{fig_island_width}). $ \mathbb{S}6.e $ is obtained from simulation $\mathbb{S}6.a$ by setting the fluctuations $\tilde{\psi}_0 = 0$.\\
One element checked in the simulations is how these variations of the zonal fields (``zonal field'' is used to refer to both the zonal current and the zonal flow) impact the maximum width (an $m=2$ island has two O-points where the separatrix doesn't necessarily have the same width, the larger is taken) of the magnetic island given by mode $\psi_2$, as shown in Fig. \ref{fig_island_width}.\\
Varying the value of the dissipative parameter $\mu_0$, i.e. comparing simulations $\mathbb{S}6$, $\mathbb{S}6.a$ and $\mathbb{S}6.b$, does not give any particulartly interesting results (the widths for these particular simulations are not shown in Fig. \ref{fig_island_width}), since the island always reaches the same width with similar dynamics. This is because the zonal flow itself is not significantly impacted by the variations of the dissipative parameter $\mu_0$ (see Fig. \ref{fig_zf_comparison}), even though these variations of $\mu_0$ span a factor $250$. In particular increasing the dissipative parameter by a factor $50$ with respect to the reference value has no effect, while a further factor $5$ shows an effect, though at that point the regime of the system is much more strongly dissipative. This is further indication that the dynamics of the zonal flow in these simulations are not strongly dissipative but rather turbulent.\\
While the complete suppression of the fluctuations for a particular mode is a much less self-consistent way of intervening in a simulation, it also allows to understand more clearly the role played by that particular mode in the dynamics.\\
For $\mathbb{S}6.c$, suppression of the poloidal flow at $t = 11200 \, \tau_A$ has no appreciable effect on the magnetic island width (maximum variation is $2.5 \%$, see Fig. \ref{fig_island_width}) for an interval of time $\Delta t \approx 1000 \tau_A \approx 33 \tau^*$, suggesting that it is the magnetic island that localizes the region of sheared flow, and the latter does not have the role of limiting the radial width of the island. This is of great interest, since zonal flows are often associated with the improvement of tokamak performance and the formation of Internal Transport Barriers (ITBs) \citep{guzdar2001zonal, diamond2005zonal, fujisawa2007experimental, dong2019nonlinear}, thus the possibility that TDMIs might aid in the drive of such structure is of interest for future devices. The dynamics of the system in $\mathbb{S}6.d$ (where suppression and change in background happen at $t = 12100 \; \tau_A$, not shown in Fig. \ref{fig_island_width}) are remarkably similar to this case, meaning that the background profile does not have a major impact on the evolution of the TDMI. Fig. \ref{fig_island_width} shows, however, that the suppression of the zonal flow halts (or at least slows down significantly) the transition from having a dominant $m=2$ mode to $m=1$ (i.e. the coalescence process).\\
For $\mathbb{S}6.e$ suppression of $\tilde{\psi}_0$ at $t = 11600 \, \tau_A$ leads to an acceleration of the coalescence process, where the $m=1$ mode grows much quicker than in other simulations (see Fig. \ref{fig_island_width}). Before this significant departure in the speed of the coalescence takes place (i.e. for $t \leq 12450 \tau_A$), the maximum difference in the island width of this case with the reference $\mathbb{S}6$ is $7.5 \%$ of the original value. Furthermore, comparing the isocontours of $\psi$ between this simulation, shown in Fig. \ref{fig_isoc_noZC} to those taken from $\mathbb{S}6$ and shown in Fig. \ref{fig_isoc_s6}, highlights the disappearance of the turbulent structures inside the larger island, and the appearance of an isolated maximum of $\psi$ off-resonance. This is coherent with what shown in Fig. \ref{fig_zonal_current}, since the zonal current creating a region of weak magnetic shear would favour the development of turbulence over a broader radial region, whereas the re-establishment of the magnetic field gradient following the suppression of the zonal current would suppress the turbulence. This also feeds back into the comparison between $\mathbb{S}4$ and $\mathbb{S}5$ in section \ref{sec_generation}, since the simulation that formed TDMIs was the one with weaker magnetic shear, that the zonal current contributes to.\\
The initial conclusion drawn from these observations is that the zonal flow is driven by the TDMI (similar observations of localized, though non-axisymmetric, flows have been reported in the literature \citep{hornsby2012dynamics, navarro2017effect, leconte2023vortex} for varied systems in the presence of magnetic islands), but its presence is needed to allow the coalescence to proceed and have energy flow into the larger scales, thus the zonal flow acts as a catalyst for the formation of TDMIs, whereas the zonal current keeps energy in the smallest scales of the system favouring the development of turbulence in the region of small magnetic shear that forms, thus inhibiting the formation of TDMIs.

\begin{table}
\begin{ruledtabular}
\begin{tabular}{|c|c|c|c|c|}
% ine
\textbf{Name} & $\bm{\mu_0}$ & $\bm{\tilde{\phi}_0}$ & $\bm{\tilde{\psi}_0}$ & $\bm{\partial_x \phi_{eq}}$ \\
$ \mathbb{S}6 $ & $5\cdot 10^{-4}$ & free & free & $- \partial_x p_{i \, eq}$ \\% ine
% \rowcolor{LightGray}
$ \mathbb{S}6.a $ & $1\cdot 10^{-5}$ & free & free & $- \partial_x p_{i \, eq}$ \\% ine
$ \mathbb{S}6.b $ & $2.5\cdot 10^{-3}$ & free & free & $- \partial_x p_{i \, eq}$ \\% ine
% \rowcolor{LightGray}
$ \mathbb{S}6.c $ & $1\cdot 10^{-5}$ & $0$ & free & $- \partial_x p_{i \, eq}$ \\% ine
$ \mathbb{S}6.d $ & $1\cdot 10^{-5}$ & $0$ & free & $0$ \\% ine
% \rowcolor{LightGray}
$ \mathbb{S}6.e $ & $1\cdot 10^{-5}$ & free & $0$ & $- \partial_x p_{i \, eq}$\\% ine
\end{tabular}
\caption{\label{tab_variations_zonal}Summary of varied parameters in the non-linear slab simulations. $\mu_0$ is the viscosity parameter applied to the mode $m=0$ of $\phi$, all other modes have the parameter $\mu$ indicated in Tab \ref{tab_typical_parameters_2D}}
\end{ruledtabular}
\end{table}

\begin{figure}
\centering
\begin{subfigure}{.45\textwidth}
\includegraphics[width = \textwidth, trim=0 0 0 3cm, clip]{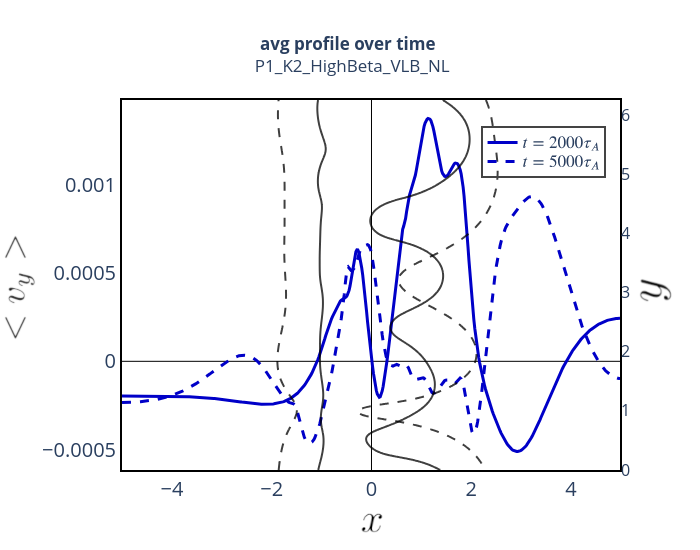}
\caption{\label{fig_zonal_flow}Behaviour of the poloidally average poloidal flow (i.e. the zonal flow). Regions of strong shear form where the island has its maximum radial width.}
\end{subfigure} \,
\begin{subfigure}{.45\textwidth}
\includegraphics[width = \textwidth, trim=0 0 0 3cm, clip]{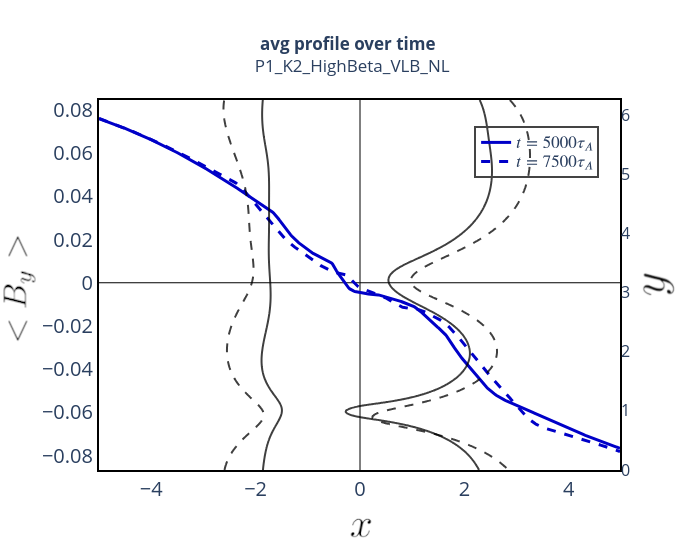}
\caption{\label{fig_zonal_current}Behaviour of the poloidally average poloidal magnetic field (i.e. the zonal current). A weakening of the magnetic shear is visible around the resonance.}
\end{subfigure}
\caption{\label{fig_zonal_fields}Evolution of the zonal fields over time compared to the phases in the evolution of the magnetic island for a non-linear fluid simulation with the same parameters as $\mathbb{S}1$ but with $L_x=10$. The blue lines are the poloidally averaged fields, with the dash of the line distinguishing different time-points. The grey lines are the isocontours of $\psi$ in the $(x,y)$ plane at times corresponding to the poloidally average profiles based on the dash of the line.}
\end{figure}

\begin{figure}
\centering
\includegraphics[width=\DefFigWidth]{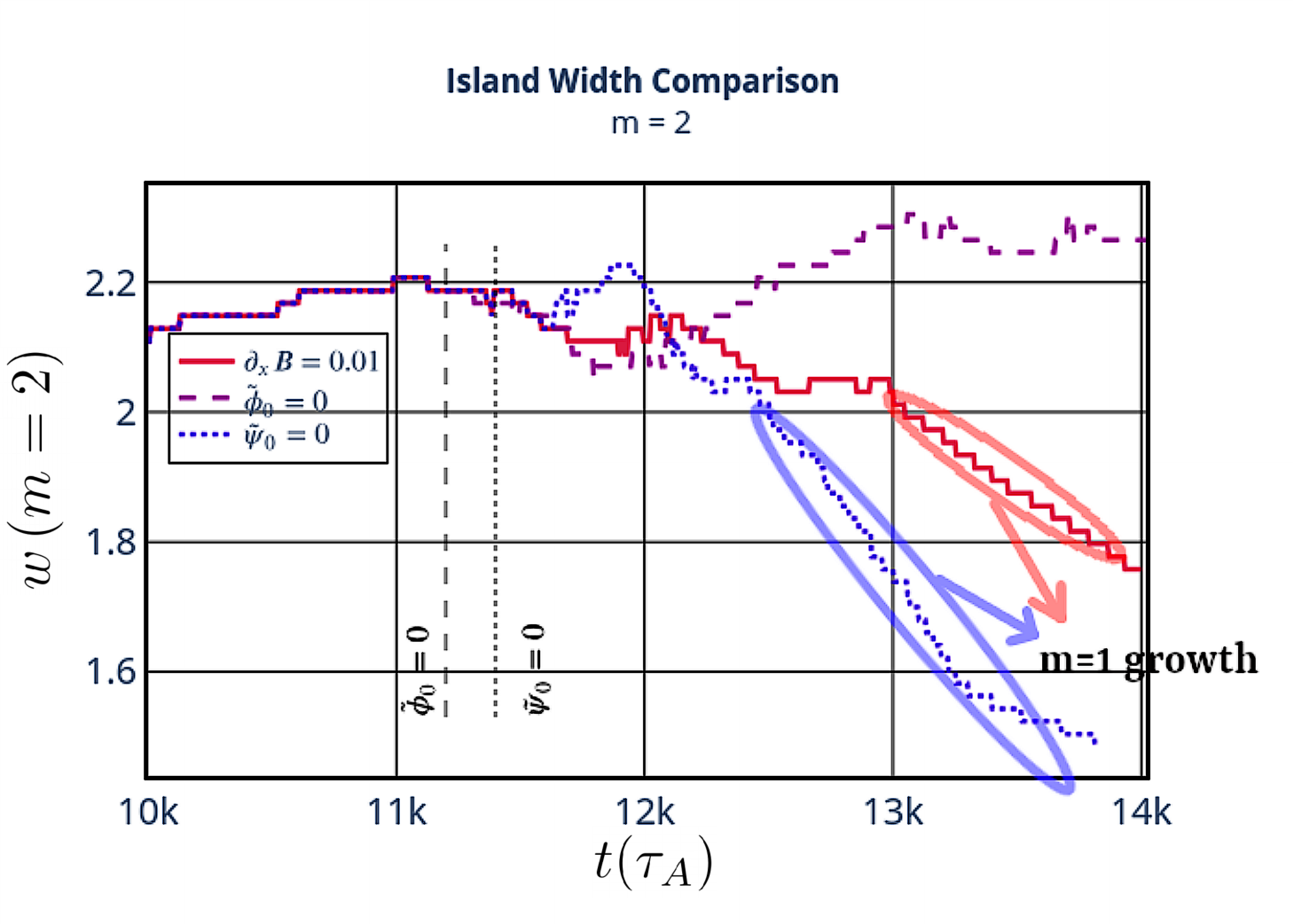}
\caption{\label{fig_island_width}Evolution of the width of the $m=2$ mode for simulations {($\mathbb{S}6 = \partial_x B_{eq} = 0.01$, $\mathbb{S}6.c = \tilde{\phi}_0 = 0$ and $\mathbb{S}6.e = \tilde{\psi}_0 = 0$)} described in Tab. \ref{tab_variations_zonal}. Varying the evolution of the zonal fields once the mode $m=2$ is well-established and dominant in the system. Note that here only the mode $m=2$ is kept in the analysis, thus the width of the structure will not match the width in the contour plots.}
\end{figure}

\begin{figure}
\centering
\includegraphics[width=\DefFigWidth, trim=0 0 0 3cm, clip ]{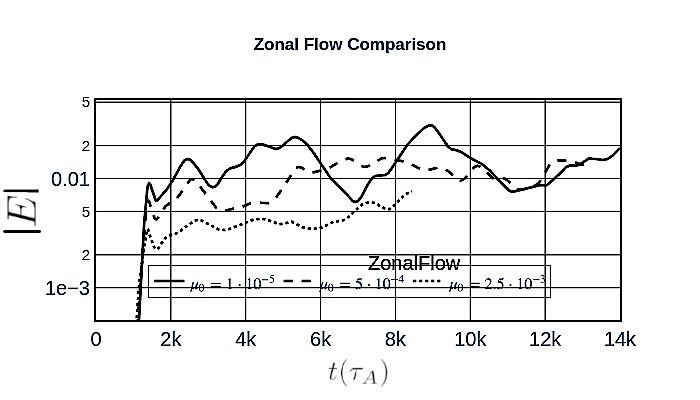}
\caption{\label{fig_zf_comparison}Evolution of the energies of the zonal flow in simulations with varying values for the dissipative parameter $\mu_0$ (simulations $\mathbb{S}6$, $\mathbb{S}6.a$ and $\mathbb{S}6.b$ in Tab. \ref{tab_variations_zonal}). The width of the magnetic island is not affected by the variation of $\mu_0$.}
\end{figure}

\begin{figure}
\centering
\begin{subfigure}{.45\textwidth}
\includegraphics[width=\textwidth, trim=0 0 0 3cm, clip]{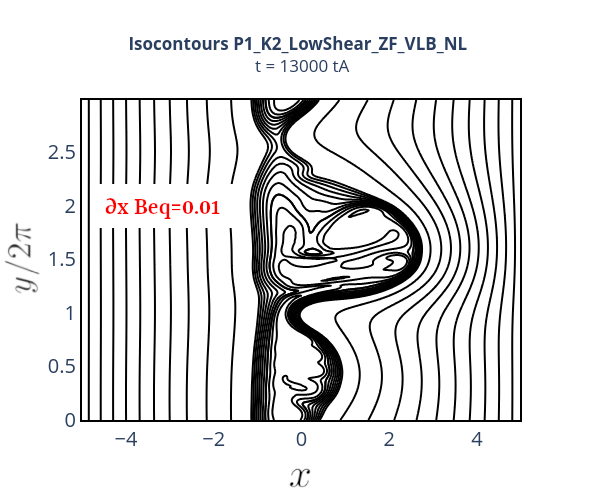}
\caption{\label{fig_isoc_s6} $\mathbb{S}6$}
\end{subfigure} \,
\begin{subfigure}{.45\textwidth}
\includegraphics[width=\textwidth, trim=0 0 0 3cm, clip]{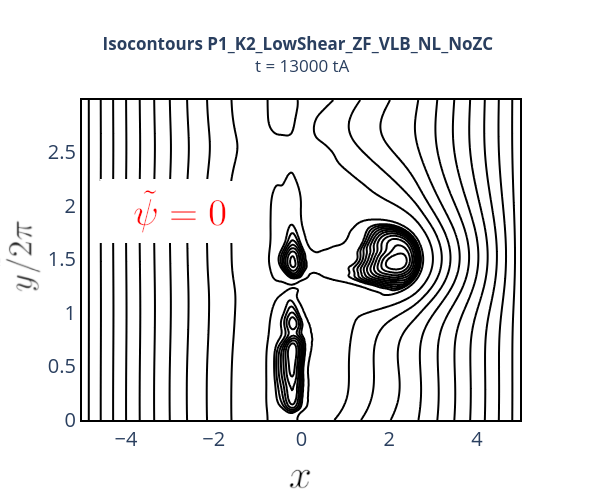}
\caption{\label{fig_isoc_noZC} $\mathbb{S}6.e$}
\end{subfigure}
\caption{Isocontours of $\psi$ at $t= 13000 \tau_A$ for simulations that run with the same dynamics up to $t = 11600 \tau_A$, where for $\mathbb{S}6.e$ the fluctuations $\tilde{\psi}_0 = 0$.}
\end{figure}

\subsection{Coupled dynamics of the interacting structures}
The very tight relation between the dynamics of the zonal current and of the magnetic island is also visible in Fig. \ref{fig_ZCMI}. Here the axes are given by the energies of two large-scale structures, and for each point the size is related to the amplitude of the turbulence and the color to the time. The energies of the modes of a generic field $f$ (actually $\psi$ and $\phi$) are computed as:
\begin{eqnarray} \label{eq_energy_mode}
\nonumber E_{f \, m} & = & (\partial_x f_m \bm{\hat{x}}, i k_y f_m \bm{\hat{y}}) \! \cdot \! (\partial_x f_m \bm{\hat{x}}, i k_y f_m \bm{\hat{y}})^* = \\
& & = (\partial_x f_m)^2 + (k_y f_m)^2
\end{eqnarray}
where in Eq. \ref{eq_energy_mode} the ``$*$'' indicates the complex conjugate, and when referring to turbulence from here forwards this indicates the sum of the energies of all modes with $m \geq 6$.\\
Because in simulation $\mathbb{S}1$ the formation of the mode $m=1$ only happens once the magnetic fluctuations are too close to the boundary of the simulation box, here ``magnetic island'' (``$MI$'') refers to $m=2$.\\
The energy dynamics thus summarized in Fig. \ref{fig_ZCMI} indicate that the energy of the magnetic island and that of the zonal current are linearly correlated, with a variation in slope (time-point $t \approx 8250 \tau_A$ highlighted in Fig. \ref{fig_ZCMI}) as the amplitude of the turbulence starts to decrease significantly. This linear correlation indicates clearly a strong interdependence between the two structures, as well as two different regimes in their interaction. Likewise, always in Fig. \ref{fig_ZCMI}, one can notice a remarkable correlation to the decrease in the magnitude of the turbulence once the zonal current starts growing more quickly, at $t \geq 8250 \tau_A$. However, the interpretation of the dynamics requires more care, and in particular the consideration of both what's shown in Fig. \ref{fig_ZCMI} and in Fig. \ref{fig_ZFMI}. For starters, this is probably not a one-way interaction, as was hinted at in the analysis of Fig. \ref{fig_island_width}, but more like a predator-prey scenario \citep{diamond2005zonal}, based on energy exchange. This is supported by the fact that, in the case considered in Fig. \ref{fig_island_width}, removing the zonal current suppressed turbulence by strengthening the magnetic shear (i.e. in that case weaker zonal current also meant weaker turbulence, while here it's stronger zonal current that's associated to weaker turbulence).  Still, the straight-forward relation between the magnetic island's and the zonal current's energy is a remarkable feature of these systems, though, it must also be remarked that such behaviour of having two phases of linear correlation for the energies of the magnetic structures is clearly identifiable only in the most unstable simulations considered in this paper, while others (including $\mathbb{S}5$) don't show such clear features, even though they form TDMIs as well.\\
Another importance piece of information in understanding the dynamics is summarized in Fig. \ref{fig_ZFMI}. Here is shown the much more complex interaction between the magnetic island, the turbulence and the zonal flow. There is an initial phase (where $E_{MI} < 0.5$) where the zonal flow grows quickly, suppressing the turbulence, as expected from usual predator-prey dynamics. This is where the change in parity of the linear modes takes place, which in turn extracts energy from the zonal flow. Once the zonal flow's energy drops to lower levels, there is a prolonged phase where the island and the turbulence grow significantly while the zonal flow remains at low energies. At the same time-point $t \approx 8250 \tau_A$ highlighted for the analysis of the zonal current (Fig. \ref{fig_ZCMI}), turbulence starts to decrease in magnitude as the zonal flow starts to grow. One can also note that the island grows at almost constant pace throughout the simulation, while the zonal flow has a strong acceleration in the late non-linear phase, possibly indicating that the zonal flow's growth is facilitated by the growth of the TDMI.\\
As a final clue to understand the dynamics, a simulation was restarted with same parameters as $\mathbb{S}6$ from $t = 12000 \tau_A$, suppressing entirely the fluctuations of all modes with $m > 5$ for all fields, and it was observed that for $\sim 200 \tau_A$ the evolution of the system in all the diagnostics traced here matches that of the reference case, indicating that the turbulence acts to extract energy from the system initially, but afterwards takes somewhat of a background role, slowly feeding the large scale modes as they interact among themselves, with a separation of time-scales.\\
Considering everything presented thus far, a path starts to appear from these results: the linearly unstable modes drive, in the very early non-linear phase, turbulence and zonal fields. The linearly unstable modes change their radial structure (parity) by subtracting energy from the zonal flow. This allows small scale modes, now with changed parity, to feed the coalescence process through the dynamics of the inverse cascade. Alongside the magnetic island, the zonal current grows as well, creating a region of weak magnetic shear where turbulence can accumulate energy, continuously extracting energy from the background. A critical point is reached where the drive of the zonal fields by the turbulence causes growth of the former at the expense of the latter, and the two zonal fields affect the turbulence in opposite ways (the sheared zonal flow suppresses turbulence, the broad zonal current at the resonance favours turbulence). Indeed, the magnetic islands are somewhat of a passive element in these complex dynamics, as the interaction between the small-scale turbulence and the zonal fields is the main effector on the dynamics, but once a magnetic island has formed, its impact on the overall system can be non-negligible (e.g. through NTMs). {It will be especially interesting to look at the behavior in 3D simulations, where the overlap of these small TDMIs can lead to the formation of stochastic regions that would disrupt the dynamics of the other structures.}\\
The simulations in this work didn't allow to observe clearly whether it's the competition between the two zonal fields that determines the saturation of the magnetic island, but this remains a possibility. Probably the lack of saturation is due to the fact that the pressure profile is never completely flattened in these gradient-driven simulations, which means that the linear drive is always present, so the coupling of a mode with itself is always possible, allowing for the unbounded growth of the zonal fields.

\begin{figure}
\centering
\includegraphics[width = \DefFigWidth, trim=0 0 0 3cm, clip]{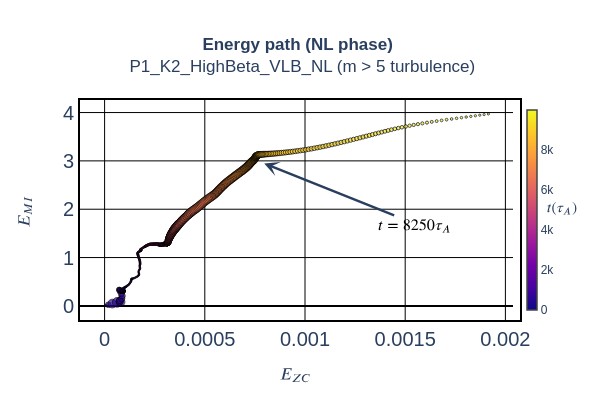}
\caption{\label{fig_ZCMI} Evolution of the energies of the zonal current (``$ZC$'') and the magnetic island (``$MI$'', corresponding to mode $m=2$). The color of the dots represents the time, while the size of the dot itself is related to the energy of the turbulence (larger dot means larger turbulence).}
\end{figure}

\begin{figure}
\centering
\includegraphics[width = \DefFigWidth, trim=0 0 0 3cm, clip]{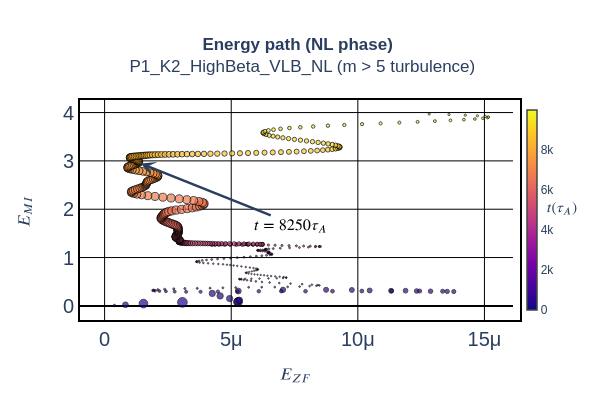}
\caption{\label{fig_ZFMI} Evolution of the energies of the zonal flow (``$ZF$'') and the magnetic island (``$MI$'', corresponding to mode $m=2$). The color of the dots represents the time, while the size of the dot itself is related to the energy of the turbulence (larger dot means larger turbulence).}
\end{figure}

\subsection{Impact on transport of TDMIs} \label{sec_transport}
In simulations where magnetic islands are present, one should expect the pressure profile to become flat within the separatrix following the formation of the inner flux-surfaces in the TDMI \citep{fitzpatrick1995helical}. However, as shown thus far in this work, TDMIs strongly interact with zonal fields, and zonal fields are known to suppress turbulence \citep{guzdar2001zonal, diamond2005zonal, fujisawa2007experimental, dong2019nonlinear}, and possibly lead to the formation of Internal Transport Barriers (ITBs) \citep{benkadda2001bursty, ida2018internal}, raising the question about the transport properties of a system that presents both kinds of structures.\\
The main results in this regard are visible in Fig. \ref{fig_press_base}, where the separatrix of the magnetic island formed by dominant $\psi$ mode at two particular times is plotted on the $(x,y)$ plane, with the poloidally averaged pressure profile superimposed. Like for the results on zonal fields (see Figs. \ref{fig_zonal_flow} and \ref{fig_zonal_current}), the simulation considered here is $\mathbb{S}1$ but with $L_x=10$.\\
The behaviour of the pressure is a combination of the behaviour expected in the presence of a magnetic island and that expected in the presence of zonal fields. At the resonance, where the instability is driven and the poloidal flow is strongly sheared (see Fig. \ref{fig_zonal_flow}) the pressure gradient is maintained almost unaltered throughout the simulation. This keeps driving turbulence while the rest of the dynamics take place. Further away from the resonance, where the magnetic island has established its magnetic flux-surfaces and the poloidal flow is not strongly sheared (see data for $t = 5000 \: \tau_A$ in Fig. \ref{fig_zonal_flow}) one can see the flattening of the pressure profile.\\
One would expect the flattening of the pressure to happen once the critical width identified in \cite{fitzpatrick1995helical} is reached. This width is shown in Fig. \ref{fig_press_base} with the arrows at the bottom of the plot. The definition for the critical width is:
\begin{equation} \label{eq_critical_island_size}
0 = \chi_\perp \Delta_{\perp} T + \chi_\parallel \Delta_{\parallel} T \; \Rightarrow \; w_c = \sqrt{8} \left( \frac{\chi_{\perp}}{\chi_{\parallel}} \right)^{\frac{1}{4}} \left( \frac{R_0 \, r_r}{n \, s_r} \right)^{\frac{1}{2}}
\end{equation}
However, it is worth noticing how the gradient at the resonance is maintained for several parallel collisional time-scales $\tau_{c \, \parallel}$, defined as
\begin{equation}
\tau_{c \, \parallel} = \frac{L^2_\parallel}{\chi_{\parallel \, i}} \approx 177 \: \tau_A
\end{equation}
after the critical island width has been reached.\\
This has the consequence that basing predictions for the impact on transport of magnetic islands on Fitzpatrick's computation might not capture entirely the role of magnetic islands in fusion plasmas, and might only apply to cases where the zonal flow is sufficiently weak. One such case would be the case of linearly unstable tearing modes, as in that case the large-scale magnetic island is directly due to the instability, and the island itself is the saturation mechanism of the instability, thus a single dominant mode determines the dynamics of the system. In the cases considered here, since the large-scale structures form as a result of dynamics at smaller scales transferring energy, their magnitudes are more closely comparable.\\
Furthermore, since one of the main experimental diagnostics for the presence of magnetic islands is the flattening of the plasma pressure \citep{snape2012influence,ida2018hysteresis,choi2021interaction}, an island that only partially flattens the pressure might be overlooked, or only be detected with a significant delay with respect to the moment it reaches the critical width $w_c$ of Eq. \ref{eq_critical_island_size}. It might be worth looking for the ``two-sided'' flattening of the pressure over smaller regions around the resonance visible in Fig. \ref{fig_press_base} for $t = 7500 \tau_A$ in experiments, to see if this can also be correlated to the presence of small islands in those regions.\\
In order to better quantify the effect of these dynamics on the confinement properties of the system, it would be much more ideal to set up flux-driven simulations, as then it is much more straightforward to establish a power balance and quantify the effect of the structures on the fluxes. In the cases presented here, the boundary values are fixed, thus fluxes are ``artifically'' set at the boundaries to ensure Dirichlet boundary conditions, though simulations were set up with buffer regions within the computational domain that delay the intervention of these artificial fluxes. It is thus reasonable to make some statements about transport in the systems analyzed here, as long as the dynamics remain separate enough from the radial boundaries of the box.\\
The peculiar transport properties of TDMIs are most noticeable when comparing Fig. \ref{fig_press_base} to Fig. \ref{fig_press_S4}. In the latter, the relaxation of the pressure profile has properties much more in line with what expected for a turbulent system, in particular the increased flux (i.e. reduced gradient) at the resonance, where the turbulence develops and persists over time. The regions of flat pressure profile visible to the sides of the resonance in Fig. \ref{fig_press_base} are completely absent in Fig. \ref{fig_press_S4}. This different behaviour can be attributed to the formation of the TDMI as the zonal fields of simulation $\mathbb{S}4$ have  similar radial structures to those shown for $\mathbb{S}1$ in Fig. \ref{fig_zonal_fields} (with maximum amplitudes slightly over $1/2$ those in the Fig. \ref{fig_zonal_fields}) and all other parameters are identical (the width of the radial box is not expected to influence the behaviour at the resonance as long as the island in $\mathbb{S}1$ does not reach excessive widths).

\begin{figure}
\centering
\begin{subfigure}{.45\textwidth}
\includegraphics[width = \textwidth, trim=0 0 0 3cm, clip]{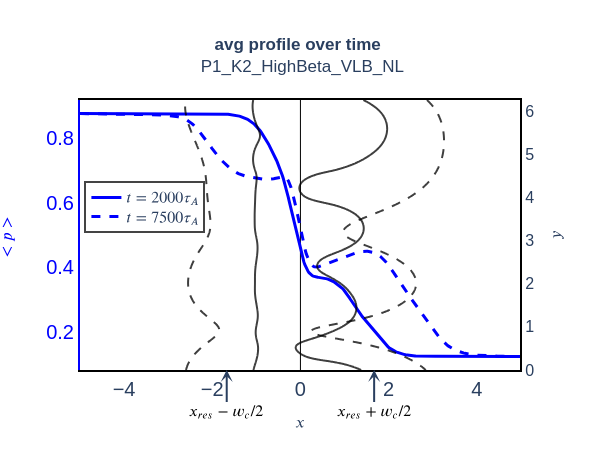}
\caption{\label{fig_press_base} Simulation with the same parameters as $\mathbb{S}1$ but with $L_x=10$. At the bottom of the figure the arrows indicate the width $w_c$ computed with Eq. \ref{eq_critical_island_size}.}
\end{subfigure} \,
\begin{subfigure}{.45\textwidth}
\includegraphics[width = \textwidth, trim=0 0 0 3cm, clip]{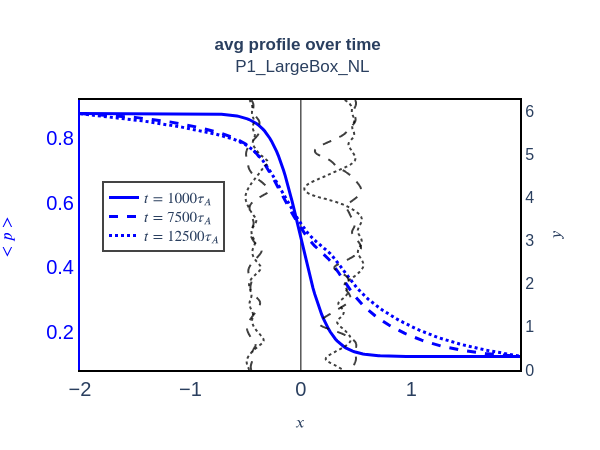}
\caption{\label{fig_press_S4} Simulation $\mathbb{S}4$. }
\end{subfigure}
\caption{Behaviour of the poloidally averaged pressure profile in non-linear fluid simulations. The blue lines are the poloidally averaged pressure, with the dash of the line distinguishing different time-points. The grey lines are the isocontours of $\psi$ (approximately the separatrix) in the $(x,y)$ plane at times corresponding to the poloidally average profiles based on the dash of the line.}
\end{figure}

%\subsection{The role of the cubic terms on the non-linear dynamics of the system} \label{sec_dynamics_cubic}

\section{Discussion} \label{sec_discussion}
The formation of magnetic islands in simulations that only exhibit linear interchange-like instabilities has already been shown to take place in the literature \citep{yagi2007nonlinear, muraglia2011generation, agullo2017nonlinearII, dubuit2021turbislands} through a mechanism of direct non-linear coupling between neighbouring modes. In this paper, it was shown how the formation of TDMIs can also happen when such direct coupling, though still present, is not sufficient to form dynamically relevant islands directly. In this case, an additional process that starts from small-scale islands forming at the linearly unstable scales and then merging to progressively larger scales through a coalescence process allows the otherwise sub-dominant islands to appear in the system. The coalescence happens on the same time-scales as the inverse non-linear cascade that takes place in simulations where no magnetic islands form (see Fig. \ref{fig_NL_spectra_nodiff}), and is thus quite slower than the direct coupling, as the latter happens in the quasi-linear phase. The difference between simulations that form magnetic islands and those that don't is that the former undergo a change in the radial structure of the magnetic potential $\psi$ at the scales of the linear instability, such that the phase of the function is more uniform over the radial coordinate (see Figs. \ref{fig_avg_parities} and \ref{fig_NL_eigenmode}). The overall parameters of the system of the system, more than the drive of the instability itself \cite{villa2025zonal}, are crucial parameter in determining whether this change in radial structure, and thus the coalescence, can take place (see Fig. \ref{fig_koalabuddies_late}).\\
The interactions of these type of TDMIs with the zonal fields were investigated. It was shown that the TDMI is capable of driving a strongly sheared poloidal flow at the position where its separatrix has its maximum radial width (see Fig. \ref{fig_zonal_flow}), and that the presence of the zonal flow is needed for the coalescence process to continue throughout the simulation (see Fig. \ref{fig_island_width}). Evidence that the zonal flow is mostly driven by the turbulence and the magnetic islands, rather than being regulated by dissipation is visible in Fig. \ref{fig_zf_comparison} and discussed in section \ref{sec_dynamics}, where it was also highlighted how this might be an unexpected positive effect of TDMIs if confirmed, since zonal flows (and zonal fields more in general) play a strong role in the establishment of ITBs \citep{guzdar2001zonal, diamond2005zonal, fujisawa2007experimental, dong2019nonlinear}. The zonal current, instead, slows down the coalescence process (see Fig. \ref{fig_island_width}) and has a clear correlation to the evolution of the magnetic island, as the energies of the two evolve with a linear relation (see Fig. \ref{fig_ZCMI}). The role played by the zonal current is that of favouring the accumulation of energy in the small-scale turbulence by creating a region of weak magnetic shear (see Fig. \ref{fig_zonal_current}), in line with the observation that, given similar linear growth rates, simulations with weaker magnetic shear form magnetic islands, while those with stronger magnetic shears don't (see Fig. \ref{fig_koalabuddies_late}).\\
Some points in relation to the impact on transport of this kind of TDMI were addressed in section \ref{sec_transport}, where it was highlighted how the flattening of the pressure profile, expected \citep{fitzpatrick1995helical} in the presence of magnetic islands, can only be partial, due to the additional effect of the zonal flow present in these simulations. This raises the possiblity that relying on pressure flattening to detect magnetic islands might leave some magnetic islands undetected if the flattening they induce is only partial.\\
Extension of the present work to 2D cylindrical systems and to 3D simulations is a natural evolution that will allow to study more complete dynamics in more realistic conditions, and is left for future work.\\

%\begin{acknowledgments}
The authors would like to thank S. Mazzi, M. J. Pueschel{, S. Cerri and E. Poli} for interesting discussions on the topic. {The anonymous referees are acknowledged for their constructive and interesting comments.}\\
The project leading to this publication has received funding from the Excellence Initiative of Aix-Marseille University - A*Midex, a French “Investissements d’Avenir” program AMX-19-IET-013.
The simulations in this article were run thanks to the support of EUROfusion and MARCONI-Fusion.
This work has been carried out within the framework of the EUROfusion Consortium, funded by the European Union via the Euratom Research and Training Programme (Grant Agreement No 101052200 — EUROfusion). Views and opinions expressed are however those of the author(s) only and do not necessarily reflect those of the European Union or the European Commission. Neither the European Union nor the European Commission can be held responsible for them. This work was partially funded by the Deutsche Forschungsgemeinschaft (DFG, German Research Foundation) under Germany´s Excellence Strategy – EXC 2094 – 390783311.\\
The authors have no conflicts to disclose.

\bibliographystyle{aipnum4-2}
\bibliography{./MasterBib.bib}

\appendix 
\section{Anti-aliasing in models with cubic terms} \label{sec_dealiasing}
As is well known, because the number of Fourier modes used to decompose the fields numerically is finite, one can incur in problems with aliasing of modes in non-linear runs, and thus de-aliasing gets applied to the arrays resulting from Fourier transforms. This concretely means that, given a simulation that uses $N_y$ points to resolve the domain in the periodic (poloidal in this work) direction, the number of Fourier modes that appear in the spectral portion of the computations is $\mathcal{F}_y = N_y/2 \pm 1$, depending on whether $N_y$ is even or odd, since the real Fourier transform is used. Normally, one reduces the number of modes kept in the computation to $2/3 \mathcal{F}_y$ (same goes for $\mathcal{F}_z$ if the simulation is three-dimensional) to avoid aliasing, but the inclusion of the cubic terms requires further care in the implementation of the code when it comes to the choice of the de-aliasing rule.\\
As is schematically shown in figure \ref{fig_dealiasing}, one starts with $\mathcal{F}$ modes after performing Fourier decomposition, and in the case of non-linear quadratic terms modes up to $2\mathcal{F}$ can in principle be driven through non-linear couplings. To see this one can write the three-wave coupling for generic functions $f$, $g$ and $h$ as:
\begin{equation}
f_m = \sum \limits_{j = 0}^\mathcal{F} \sum \limits_{k = 0}^\mathcal{F} g_j h_{k} \; \delta (m - j + k)
\end{equation}
where $\delta (m - j + k)$ is the Kronecker $\delta$. Since both $j$ and $k$ can go up to $\mathcal{F}$, the maximum value of $m$ is, in principle, $m = 2\mathcal{F}$.\\
Since $\mathcal{F}$ is finite, however, this means that any mode $m = \mathcal{F}+k$ driven by non-linear couplings gets aliased as the mode $ m' = k-2\mathcal{F}-1$. In terms of mode number one has the following aliasing for all $m = \mathcal{F}+k > \mathcal{F}$:
\begin{equation}
\mathcal{F}+k \Rightarrow k - \mathcal{F} - 1
\end{equation}
One then proceeds to look for a number of modes $M_q < \mathcal{F}$ that leaves a range in the mode-numbers where amplitudes can be put to $0$, such that no aliasing shows up in the system. For quadratic terms the condition one looks for is that the maximum non-linear mode that can be driven, $2M_q$, falls within the de-aliasing range. This means (in terms of the indeces of the Fourier transform array):
\begin{equation}
2M_q \leq (\mathcal{F}-M_q) + \mathcal{F}
\end{equation}
\begin{equation}
M_q \leq \frac{2}{3}\mathcal{F}
\end{equation}
However, for cubic terms, the maximum mode that can be driven is $3M_c$, such that the de-aliasing rule should be:
\begin{equation}
3M_c \leq (\mathcal{F}-M_c) + \mathcal{F}
\end{equation}
\begin{equation}
M_c \leq \frac{1}{2}\mathcal{F}
\end{equation}
This is immediately generalized for a given degree of non-linearity $\lambda$ as
\begin{equation}
M_\lambda \leq \frac{2}{\lambda + 1}\mathcal{F}
\end{equation}
This de-aliasing rule has been applied in this work for all simulations that used the cubic terms.

\begin{figure}[H]
\centering
\includegraphics[width=\DefFigWidth]{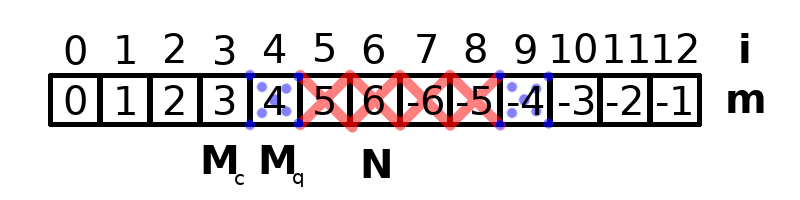}
\caption{\label{fig_dealiasing}Schematic representation of the de-aliasing process. The numbers in the squares (row ``m") indicate the mode numbers as they are stored in an array resulting from Fast Fourier Transform (FFT). The numbers above the squares (row ``i") indicate the indeces of the modes in the array. The red crosses indicate the slots that get set to $0$ in quadratic de-aliasing. The blue dotted crosses indicate the additional slots that should be set to zero for de-aliasing with cubic terms.}
\end{figure}

\end{document}